\documentclass[12pt,aps,prd,tightenlines,nofootinbib,floatfix]{revtex4-2}
\usepackage{latexsym,amsmath,amssymb,amsbsy,graphicx}
\usepackage{xcolor}

\begin{document}
	
\title{Exclusive semileptonic decays of $D$ and $D_s$ mesons into  orbitally and radially excited states of strange and light mesons}

\author{V.\,O. \surname{Galkin$^1$}} 
\email{galkin@ccas.ru}
\author{I.\,S. \surname{Sukhanov$^{1,2}$}}
\email{sukhanov.is17@physics.msu.ru}

\affiliation{$^1$
	Federal Research Center ``Computer Science and Control'' Russian Academy of Sciences, Vavilov Street 40, Moscow, 119333 Russia}
\affiliation{$^2$ 
 Faculty of Physics, M.V. Lomonosov Moscow State University, Leninskie Gory 1-2,
	Moscow 119991, Russia.}


\preprint{}	
	
\begin{abstract}
Semileptonic decays of $D$ and $D_s$ mesons into orbitally and radially excited strange and light mesons are studied in detail within the framework of the relativistic quark model based on the quasipotential approach and quantum chromodynamics. The hadronic matrix elements of the weak current between meson states are calculated with the consistent account of relativistic effects including contributions of the intermediate negative energy states and boosts of the meson wave functions from the rest to moving reference frame. The invariant form factors  that parameterize these matrix elements are obtained as the overlap integrals of the initial and final meson wave functions. Their dependence on the square of the transferred momentum $q^2$ is explicitly determined within the whole accessible kinematic range. A convenient analytical approximation for numerical values of form factors is given.  These form factors and helicity formalism are employed for the calculation of the differential and total semileptonic decay rates of charm mesons  into excited strange and light mesons.  Different asymmetry and polarization parameters are also evaluated. The obtained results are compared  with other theoretical calculations and available experimental data. Reasonable agreement with experimentally measured semileptonic decay branching fractions and upper bounds is obtained. 
\end{abstract}

\maketitle

\section{INTRODUCTION} \label{sec:intro}

Semileptonic decays of heavy mesons provide an important information on the values of the Cabibbo-Kobayashi-Maskawa (CKM) matrix elements $V_{Qq}$ (with $Q$ denoting the heavy quark and $q$ the light one), which are essential parts of the standard model. Experimentally such decays can be measured more accurately than the pure leptonic ones since there is no helicity suppression for them and thus semileptonic decays have substantially larger branching fractions. Theoretically semileptonic decays occupy intermediate position between pure leptonic and nonleptonic decays,  being more difficult to study than pure leptonic ones, but significantly less complicated than hadronic ones, since they contain one meson and a lepton pair in the final state. The lepton part is easily calculated using standard methods, while the hadronic part factorizes thus reducing theoretical uncertainties. The hadronic matrix element is usually parameterized by the set of invariant form factors, which are calculated using nonperturbative approaches based on quantum chromodynamics (QCD), such as lattice QCD, QCD sum rules, potential quark models.

At present essential experimental progress has been achieved in studding exclusive semileptonic decays of charmed mesons \cite{ParticleDataGroup:2024cfk}. Many semileptonic decay channels both of the $D$ and $D_s$ mesons to the ground states of strange and light mesons were precisely measured by the BESIII Collaboration \cite{Ablikim:2020tmg,Ablikim:2020hsc,BESIII:2021pvy,BESIII:2023gbn,BESIII:2024lnh,BESIII:2024mot,BESIII:2024zft,BESIII:2024slx,BESIII:2024njj,BESIII:2024xjf}. These studies allowed not only to determine the semileptonic decay branching fractions but also to evaluate some of the decay form factors at the zero recoil point of the final meson $q^2=0$ \cite{Ablikim:2020hsc,BESIII:2023gbn,BESIII:2024mot,BESIII:2024zft,BESIII:2024slx,BESIII:2024njj,BESIII:2024xjf} and the forward-backward asymmetries \cite{BESIII:2023gbn}. The newly obtained data is in agreement with our previous studies in the framework of the relativistic quark model of the charmed meson semileptonic decays to the ground state strange and light mesons \cite{Faustov:2019mqr}. Note that the two-body nonleptonic heavy meson decays were also studied in this model within the factorization approach~\cite{Yu:2022ngu}. 

Recently first experimental data on the semileptonic $D_{(s)}$ decays to orbitally excited strange and light mesons became available \cite{CLEO:2007oer,BESIII:2019eao,BESIII:2020jan,BESIII:2021uqr,BESIII:2023clm,BESIII:2024pwp,BESIII:2025yot}. The branching fractions of the semielectronic  $D$ decays into the axial-vector $K_1(1270)$ meson were measured in Refs.~ \cite{BESIII:2019eao,BESIII:2021uqr} and very recently semimuonic ones were reported in Ref.~\cite{BESIII:2025yot}. A rather strict upper bounds for the branching fractions of semileptonic $D$  decays to the axial-vector $b_1(1235)$ \cite{BESIII:2020jan} and $D_s$ to  $K_1(1270)$ and $b_1(1235)$ were set \cite{BESIII:2023clm}. Very recently the products of the branching fractions $Br(D^0\to b_1(1235)^-e^+\nu_e)\times Br(b_1(1235)^-\to \omega\pi^-)$ and $Br(D^+\to b_1(1235)^0e^+\nu_e)\times Br(b_1(1235)^0\to \omega\pi^0)$ have been reported by the BESIII Collaboration \cite{BESIII:2024pwp}. Current and future experiments at charm-tau factories are expected to yield new, more complete and accurate data on the semileptonic decays of $D$ and $D_{s}$ mesons to excited strange and light mesons.

Theoretically semileptonic decays of charmed mesons to excited charmed and light mesons are significantly less studied than decays to the ground states. An additional complication results from the ambiguous interpretation of the orbitally excited light and strange meson states. For example, there is still no consensus  about interpretation of the light scalar meson states with masses below 1~GeV \cite{ParticleDataGroup:2024cfk}. However, most of theoretical approaches (including our model \cite{Ebert2009}) predict the first orbital excitations above 1~GeV and interpret the lower scalar states as tetraquark or molecular states \cite{Jaffe:2004ph,Ebert:2008id}. The other complication arises from the probable admixture of the glueball to the isospin singlet states \cite{Cheng:2015iaa,Klempt:2021wpg}.  In literature mostly semileptonic decays to axial-vector and scalar mesons were considered within covariant light-front quark model \cite{Cheng2017},  QCD and light-cone sum rules \cite{Zuo2016,Khosravi2009,Momeni2019,Huang2021,Huang2023,Yang2024,Yang2006,Hu2022}, Hard-Wall AdS/QCD model \cite{Momeni2022} and SU(3) flavor symmetry \cite{Wang:2022yyn,Qiao2024}. The semileptonic $D_{(s)}$ decays to tensor states were only studied in the SU(3) flavor analysis in Ref.~\cite{Qiao2024}.

In this paper we calculate the matrix elements of the flavor changing charged weak current between the initial $D$ or $D_s$ mesons and final strange or light  orbitally excited scalar, axial-vector and tensor mesons  as well as radially excited pseudoscalar or vector mesons in the framework of the relativistic quark model based on the quasipotential approach. On this basis we determine the corresponding decay form factors. Note that the employed model has been successfully applied for the calculations of the hadron spectroscopy \cite{Ebert2003} and weak decays \cite{Faustov1995,Ebert2003a,Faustov:2012mt,Faustov:2022ybm}. It was found that relativistic effects play a very important role both for light and heavy hadrons. Thus form factors are calculated with the consistent account of the relativistic quark dynamics. They are expressed through the overlap integrals of the meson wave functions which are known from the study of their spectroscopy. The momentum transfer squared $q^2$ dependence of form factors is explicitly determined in the whole kinematical range without additional assumptions and extrapolations. Then we use these form factors and the helicity formalism for the calculation of the differential and total branching fractions as well as polarization and asymmetry parameters.  We  compare our results with available experimental data \cite{ParticleDataGroup:2024cfk,CLEO:2007oer,BESIII:2019eao,BESIII:2020jan,BESIII:2021uqr,BESIII:2023clm,BESIII:2024pwp} and previous predictions \cite{Cheng2017,Qiao2024,Khosravi2009,Momeni2019,Yang2006,Zuo2016,Yang2024,Huang2023,Huang2021,Qiao2024,Hu2022}.
Note that we present the first detailed dynamic calculation of the semileptonic decays of $D$ and $D_s$ mesons into all orbiatally excited $P$-wave, as well as radially excited $2S$-wave strange and light mesons. The sum over these channels gives an estimate of the resonance contributions to the exclusive semileptonic decay rates of $D$ and $D_s$ mesons in the energy range from 1~GeV to 1.5~GeV. The contributions of higher resonances are strongly suppressed by the small phase space. Moreover, the comparison of the obtained predictions with the available and future data can help in determining the CKM matrix elements ($V_{cs}$, $V_{cd}$) and even in revealing the nature of excited mesons in this energy range.

The paper is organized as follows. In Sec.~\ref{sec:rqm} we briefly describe our relativistic quark model. Discussion of the masses and wave functions of orbitally and radially excited light and strange mesons is given in Sec.~\ref{sec:mass}. The mixing of the isoscalar states is considered. In Sec.~\ref{sec:weak} the calculation of the weak decay matrix elements between meson states with the account of relativistic effects in the framework of the quasipotential approach is briefly summarized. The semileptonic decay form factors  and their convenient analytic parameterization which accurately reproduces the numerical results for the momentum transfer squared $q^2$ dependence of the form factors in the whole accessible kinematical range is presented in Sec.~\ref{sec:ff}. In Sec.~\ref{sec:sem} expressions  for the differential decay rates and other observables are given in terms of these form factors on the basis of the helicity formalism.  Then in Sec.~\ref{sec:results} we use the form factors to calculate the differential and total $D$ and $D_s$ meson semileptonic decay rates and different asymmetries and polarization parameters. Decays both with positrons and muons are considered. This allows us to give predictions for the ratios of the corresponding decay rates which can be used for the test of the lepton universality in charm meson decays. The detailed comparison of the obtained results with the available experimental data and previous calculations is presented. Finally, Sec.~\ref{sec:concl} contains our conclusions.

\section{RELATIVISTIC QUARK MODEL}	\label{sec:rqm}
	
For the calculation of meson properties we employ the relativistic quark model based on the quasipotential approach. In this model a meson with the mass $M$ is described by the wave function $\Psi_M(\textbf{p})$ of the quark-antiquark bound state which satisfies the Schr\"{o}dinger-like quasipotential equation \cite{Ebert2003}	
\begin{equation}\label{1}
\Bigg(\dfrac{b^2(M)}{2\mu_R}-\dfrac{\textbf{p}^2}{2\mu_R}\Bigg)\Psi_M(\textbf{p})=\int \dfrac{d^3q}{2\pi^3}V(\textbf{p},\textbf{q};M)\Psi_M(\textbf{q}),
\end{equation}
where $\textbf{p}$ is the relative quark momentum. The relative momentum squared in the center of mass system on the mass shell is given by
\begin{equation}\label{2}
b^2(M) = \dfrac{\big[ M^2-(m_1+m_2)^2\big]\big[M^2-(m_1-m_2)^2\big]}{4M^2},
\end{equation}
and the relativistic reduced mass is defined by
\begin{equation}\label{3}
\mu_R = \dfrac{M^4-(m^2_1-m^2_2)^2}{4M^3},
\end{equation}
where $m_{1,2}$ are the quark masses.

The kernel of this equation $V(\textbf{p},\textbf{q};M)$ is the QCD-motivated quark-antiquark potential which is constructed by the off-mass-shell scattering amplitude projected on the positive energy states. We assume \cite{Ebert2003} that it consists from the one-gluon exchange term which dominates at small distances and a mixture of the scalar and vector linear confining interactions which dominates at large distances. Moreover, we assume that the long-range vertex of the confining vector interaction contains additional Pauli term. Then the quasipotential is given by	
\begin{equation}\label{Ker}
V(\textbf{p},\textbf{q};M)=\bar u_1(\textbf{p)} \bar u_2(-\textbf{p})\mathcal{V}(\textbf{p},\textbf{q};M)u_1(\textbf{q})u_2(-\textbf{q}),
\end{equation}
with
\begin{equation}
\mathcal{V}(\textbf{p},\textbf{q};M)=\frac{4}{3}\alpha_s D_{\mu \nu}(\textbf{k})\gamma^\mu_1\gamma^\nu_2+V^V_{conf}(\textbf{k})\Gamma^\mu_1(\textbf{k})\Gamma_{2;\mu}(\textbf{k})+V^S_{conf}(\textbf{k}), 
\end{equation}
where $\alpha_s$ is the QCD coupling constant, $D_{\mu\nu}$ is the gluon propagator in the Coulomb gauge, $\textbf{k}=\textbf{p}-\textbf{q}$, and $\gamma_\mu$ and $u({\bf p})$ are the Dirac matrices and spinors, respectively. The long-range vector vertex has the form
\begin{equation}
\Gamma_\mu(\textbf{k})=\gamma_\mu+\frac{i\kappa}{2m}\sigma_{\mu \nu}k^\nu,
\end{equation}
where $\kappa$ is the long-range anomalous chromomagnetic quark moment. In the nonrelativistic
limit confining vector and scalar potentials reduce to
\begin{equation}
V^V_{conf}(r)=(1-\varepsilon)(Ar+B), \qquad V^S_{conf}(r)=\varepsilon(Ar+B),
\end{equation}
and in the sum they reproduce the linear potential
\begin{equation}
V_{conf}(r)=V^V_{conf}(r)+V^S_{conf}(r)=Ar+B,
\end{equation}
where $\varepsilon$ is the mixing coefficient. Thus this quasipotential can be viewed as the relativistic generalization of the nonrelativistic Cornell potential
\begin{equation}
V_{NR}(r)=-\frac{4}{3}\frac{\alpha_s}{r}+Ar+B.
\end{equation}
This quasipotential contains both spin-independent and spin-dependent relativistic contributions.

All parameters of the model were fixed from the previous consideration of hadron spectroscopy and decays \cite{Ebert2003}. Thus the  constituent quarks have the following values of masses $m_c = 1.55$~GeV, $m_s =0.5$~GeV, $m_{u,d} = 0.33$~GeV. The parameters of the linear potential are $A = 0.18$~GeV$^2$ and $B = -0.30$~GeV. 
The value of the mixing parameter of the scalar and vector confining potentials $\varepsilon=-1$ was fixed by analyzing the semileptonic decays of $B\rightarrow D$ \cite{Faustov1995}, and the radiative decays of charmonium \cite{Ebert2003}. The universal Pauli interaction constant $\kappa=-1$ was determined based on the analysis of the fine splitting of the $^3P_J$-state of heavy quarkonia~\cite{Ebert2003}.
We take the running QCD coupling constant with infrared freezing
\begin{equation}
\alpha_s(\mu)=\dfrac{4\pi}{\left( 11 - \frac23 n_f\right)\ln\dfrac{\mu^2+M^2_0}{\Lambda^2}},
\end{equation}
where  $n_f$ is the number of flavors, $\Lambda = 413$~MeV, $M_0 = 2.24\sqrt{A} = 0.95$~GeV$^3$ and the scale $\mu$ is set to twice the reduced mass of the constituents $\mu=\frac{2m_1m_2}{m_1+m_2}$.

\section{Masses and wave functions of excited light and strange mesons} \label{sec:mass}

The spectroscopy of heavy-light and light mesons was discussed in detail in Refs. \cite{Ebert2009,Ebert2010a}. The calculated masses of both ground and excited states were found in agreement with available experimental data and exhibit linear Regge trajectories. We give the calculated masses of the $1P$ and $2S$ states of light and strange mesons in Tables~\ref{Mass1}--\ref{Mass2}. Reasonable agreement between theoretical values \cite{Ebert2009} and experimental data \cite{ParticleDataGroup:2024cfk} is observed. Let us emphasize that all calculated scalar ($1^3P_0$) meson masses have values larger than 1~GeV. Scalars with masses less than 1~GeV are considered to be tetraquarks in our model \cite{Ebert:2008id}. Note that the mass of the observed scalar isotriplet ($I=1$) $a_0(1450)$ meson is more than $6\sigma$ higher than our prediction in $q\bar q$ scenario ($q=u,d$), while in the tetraquark picture with $(qs)(\bar q\bar s)$ composition the ground scalar state consisting of an axial-vector diaquark and axial-vector antidiquark has the mass in agreement with the experimental value \cite{Ebert:2008id}. Nevertheless in this paper we consider  $a_0(1450)$ to be the $1^3P_0$ meson.  In the following we use the experimental values of these masses for calculating the semileptonic decays. 

\begin{table}
	\caption{Calculated and experimental masses of excited light mesons (in MeV).}\label{Mass1}
	\begin{ruledtabular}
		\begin{tabular}{ccc@{}c@{}ccccc@{}c@{}c}
		&&&\multicolumn{2}{c}{Mass}&&\multicolumn{2}{c}{Mass}&&\multicolumn{2}{c}{Mass}\\
		\cline{4-5} \cline{7-8} \cline{10-11}
		State&$J^{PC}$&$I=1$&Th.\cite{Ebert2009}&Exp.\cite{ParticleDataGroup:2024cfk}&$I=0$&Th.\cite{Ebert2009}&Exp.\cite{ParticleDataGroup:2024cfk}&$I=0$&Th.\cite{Ebert2009}&Exp.\cite{ParticleDataGroup:2024cfk}\\
		\hline
		$1^3P_0$&$0^{++}$&$a_0(1450)$&$1176$&$1439(34)$&$f_0(1370)$&$1260$&$\{1200..1500\}$&$f_0(1500)$&$1482$&$1522(25)$\\
		$1^3P_1$&$1^{++}$&$a_1(1260)$&$1254$&$1230(40)$&$f_1(1285)$&$1281$&$1281.8(5)$&$f_1(1420)$&$1440$&$1428.4(^{1.5}_{1.3})$\\
		$1^3P_2$&$2^{++}$&$a_2(1320)$&$1317$&$1318.2(6)$&$f_2(1270)$&$1322$&$1275.4(8)$&$f_2^\prime(1525)$&$1524$&$1517.3(2.4)$\\
		$1^3P_1$&$1^{+-}$&$b_1(1235)$&$1258$&$1229.5(3.2)$&$h_1(1170)$&$1258$&$1166(6)$&$h_1(1415)$&$1484$&$1409(^9_8)$\\
		$2^1S_0$&$0^{-+}$&$\pi(1300)$&$1292$&$1300(100)$&$\eta(1295)$&$1292$&$1294(4)$&$\eta(1475)$&$1536$&$1476(4)$\\
		$2^3S_1$&$1^{--}$&$\rho(1450)$&$1486$&$1465(25)$&$\omega(1420)$&$1486$&$1410(60)$&$\phi(1680)$&$1698$&$1680(20)$\\
		\end{tabular}
		\end{ruledtabular}
\end{table}		

\begin{table}
	\caption{Calculated and experimental masses of excited strange mesons (in MeV).}\label{Mass2}
	\begin{ruledtabular}
		\begin{tabular}{c c c c c }
		&&&\multicolumn{2}{c}{Mass}\\
		\cline{4-5}
		State&$J^{PC}$&$I=1/2$&Th. \cite{Ebert2009}&Exp. \cite{ParticleDataGroup:2024cfk}\\
		\hline
		$1^3P_0$&$0^+$&$K^*_0(1430)$&$1362$&$1425(50)$\\
		$1^3P_2$&$2^+$&$K^*_2(1430)$&$1424$&$1432.4(1.3)$\\
		$1P_1$&$1^+$&$K_1(1270)$&$1294$&$1253(7)$\\
		$1P_1$&$1^+$&$K_1(1400)$&$1412$&$1403(7)$\\
		$2^1S_0$&$0^-$&$K(1460)$&$1538$&$1482(19)$\\
		$2^3S_1$&$1^-$&$K^*(1680)$&$1675$&$1718(18)$\\
		\end{tabular}
		\end{ruledtabular}
\end{table}
 The meson wave functions were also obtained during their mass spectra calculations and can be used for the evaluation of the meson decays. It is important to take into account the mixing of the $q\bar q\equiv (u\bar u+d\bar d)/\sqrt2$ and $s\bar s$ states for the isoscalar ($I=0$) mesons. 
 
 Theoretical analysis of such mixing for the tensor $f_2(1270)$ and $f'_2(1525)$ states indicate that these states are almost pure $q\bar q$ and $s\bar s$ states. Indeed, the detailed study of such mixing \cite{Cheng:2011fk,Li:2000zb,Li:2018lbd} results in the following structure of these tensor states
 \begin{eqnarray} \label{f2}
 f_2(1270)&=&f_2\!\left(\frac{u\bar u+d\bar d}{\sqrt2}\right)\cos\theta_{f_2}+f_2(s\bar s)\sin\theta_{f_2},\cr
 f'_2(1525)&=&f_2\!\left(\frac{u\bar u+d\bar d}{\sqrt2}\right)\sin\theta_{f_2}-f_2(s\bar s)\cos\theta_{f_2},
 \end{eqnarray}
 with the mixing angle $\theta_{f_2}$ about $(9\pm 1)^{\circ}$.
 
 The mixing of the axial-vector meson states is in detail discussed in Refs.~\cite{Cheng:2011pb,Cheng:2013cwa} and is given by  
 
 a) for $f_1(1285)$ and $f_1(1420)$ 
 \begin{eqnarray}\label{f1}
 f_1(1285)&=&f_1\!\left(\frac{u\bar u+d\bar d}{\sqrt2}\right)\cos\theta_{f_1}+f_1(s\bar s)\sin\theta_{f_1},\cr
 f_1(1420)&=&f_1\!\left(\frac{u\bar u+d\bar d}{\sqrt2}\right)\sin\theta_{f_1}-f_1(s\bar s)\cos\theta_{f_1},
 \end{eqnarray}

b) for $h_1(1170)$ and $h_1(1380)$  
\begin{eqnarray}\label{h1}
 h_1(1170)&=&h_1\!\left(\frac{u\bar u+d\bar d}{\sqrt2}\right)\cos\theta_{h_1}+h_1(s\bar s)\sin\theta_{h_1},\cr
 h_1(1415)&=&h_1\!\left(\frac{u\bar u+d\bar d}{\sqrt2}\right)\sin\theta_{h_1}-h_1(s\bar s)\cos\theta_{h_1},
 \end{eqnarray}
with  $\theta_{f_1}=20.3^{\circ}$ and $\theta_{h_1}=3.3^{\circ}$, respectively.

Theoretical interpretation of the mixing of the scalar $f_0(1370)$ and $f_0(1500)$ mesons is more uncertain. Usually, an additional scalar  $f_0(1710)$ meson is included into analysis. It is believed that one of these states should be predominantly a glueball \cite{Cheng:2006hu,Cheng:2015iaa,Petrov:2022ipv}. The analysis of Refs.~\cite{Cheng:2006hu,Cheng:2015iaa} concludes that $f_0(1710)$ contains the largest glueball component, while $f_0(1500)$ is predominantly flavor octet and the mixing of these states is given by
 \begin{eqnarray}\label{eq:f0}
\begin{pmatrix} f_0(1370) \\ f_0(1500) \\
f_0(1710)\end{pmatrix}=\begin{pmatrix} 0.78(2) & 0.52(3) & -0.36(1)
\\ -0.55(3) & 0.84(2) & 0.03(2) \\  0.31(1) & 0.17(1) & 0.934(4)  \end{pmatrix}
\begin{pmatrix}f_{0}\!\left(\frac{u\bar u+d\bar d}{\sqrt2}\right) \\ f_{0}(s\bar s) \\G \end{pmatrix},
\end{eqnarray}
where $G$ denotes a glueball. We adopt this mixing scheme in our following calculations.

Note that theoretical values  of the isoscalar state masses in Table~\ref{Mass1} were obtained with the account of the mixing discussed above. 

The calculated masses of the radially excited $2S$ spin-singlet $\pi(1300)$ and $\eta(1295)$ as well as spin-triplet $\rho(1450)$ and $\omega(1420)$ mesons, considered as $q\bar q$ ($q=u,d$) states, and the calculated masses of respective $\eta(1475)$ and $\phi(1680)$, considered as $s\bar s$,   agree well with the experimental values. Thus we neglect a very small mixing among them in our following calculations of the semileptonic decays.
 
 The axial vector strange meson states $K_1(1270)$  and $K_1(1400)$
are the mixtures of spin-triplet ($^3P_1$)  and spin-singlet ($^1P_1$)
states \cite{Ebert2009}:
\begin{eqnarray}
  \label{eq:mix}
  K_1(1270)&=&K(^1P_1)\cos\varphi+K(^3P_1)\sin\varphi, \cr
 K_1(1400)&=&-K(^1P_1)\sin\varphi+K(^3P_1)\cos\varphi, 
\end{eqnarray}
where $\varphi$ is a mixing angle. Such mixing occurs due to the non-diagonal spin-orbit and tensor terms in the quasipotential.  The found value of the mixing angle is
$\varphi=43.8^\circ$ \cite{Ebert2009}.

The attribution of the radially excited $2S$ states of strange mesons to the experimentally observed states is  rather uncertain. In our following calculations we assume that $K(1460)$ is the $2^1S_0$ state and $K^*(1680)$ is the pure $2^3S_1$ state, while it can also be either the $1^3D_1$ state or the mixture of these states.

 \section{Weak decay matrix elements} \label{sec:weak}
 
For the consideration of the $D$ meson semileptonic decays it is necessary to calculate the hadronic matrix element of the local current governing the $c\to f (f = s, d)$ weak transition. In the quasipotential approach the matrix element of this weak current $J^W_\mu=\overline{f}\gamma_\mu(1-\gamma_5)c$ between the initial $D_{(s)}$ meson with four-momentum $p_{D_{(s)}}$ and the final meson $F$ with four-momentum $p_F$ is given by \cite{Ebert2003a}
\begin{equation}\label{Matrix}
\langle{F(p_F)|J^W_\mu|D_{(s)}(p_{D_{(s)}})}\rangle = \int \dfrac{d^3p d^3q}{(2\pi)^6} \overline{\Psi}_{F\textbf{p}_F}(\textbf{p})\Gamma_\mu(\textbf{p},\textbf{q})\Psi_{D_{(s)}\textbf{p}_{D_{(s)}}}(\textbf{q}),
\end{equation}
where $\Psi_{M\textbf{p}_M}$ are the initial and final meson wave functions projected on the positive energy states and boosted to the moving reference frame with the three-momentum $\textbf{p}_M$. The vertex function
\begin{equation}\label{12}
\Gamma=\Gamma^{(1)}+\Gamma^{(2)},
\end{equation}
where $\Gamma^{(1)}$ is the leading-order vertex function which corresponds to the impulse approximation
\begin{equation}\label{13}
\Gamma^{(1)}_\mu(\textbf{p},\textbf{q})=\overline{u}_f(\textbf{p}_f)\gamma_\mu(1-\gamma_5)u_c(\textbf{q}_c)(2\pi)^3\delta(\textbf{p}_q-\textbf{q}_q),
\end{equation}
and contains the $\delta$ function responsible for the momentum conservation on the spectator $q$ antiquark line. The vertex function $\Gamma^{(2)}$ takes into account interaction of the active quarks $(c, f)$ with the spectator antiquark $(q)$ and includes the negative-energy part of the active quark propagator. It is the consequence of the projection on the positive energy states in the quasipotential approach and is given by
\begin{eqnarray}\label{14}
\Gamma^{(2)}_\mu(\textbf{p},\textbf{q})&=&\overline{u}_f(\textbf{p}_f)\overline{u}_q(\textbf{p}_q)\Bigg\{\mathcal{V}(\textbf{p}_q-\textbf{q}_q)\dfrac{\Lambda^{(-)}_f(\textbf{k}^\prime)}{\epsilon_f(\textbf{k}^\prime)+\epsilon_f(\textbf{q}_c)}\gamma^0_1\gamma_{1\mu}(1-\gamma^5_1)\nonumber\\&&\qquad\qquad\qquad\quad+\gamma_{1\mu}(1-\gamma^5_1)\dfrac{\Lambda^{(-)}_c(\textbf{k})}{\epsilon_c(\textbf{k})+\epsilon_c(\textbf{p}_f)}\gamma^0_1\mathcal{V}(\textbf{p}_q-\textbf{q}_q) \Bigg\} u_c(\textbf{q}_c)u_q(\textbf{q}_q),\qquad
\end{eqnarray}
where $\textbf{k}=\textbf{p}_f-\boldsymbol{\Delta};\ \textbf{k}^\prime=\textbf{q}_c+\boldsymbol{\Delta};\ \boldsymbol{\Delta}=\textbf{p}_F-\textbf{p}_D;\ \epsilon(\textbf{p})=\sqrt{\textbf{p}^2+m^2};$ and the projection operator on the negative-energy states
\begin{equation*}
\Lambda^{(-)}(\textbf{p})=\dfrac{\epsilon(\textbf{p})-(m\gamma^0+\gamma^0(\boldsymbol{\gamma}\cdot\textbf{p}))}{2\epsilon(\textbf{p})}.
\end{equation*}
Note that the $\delta$ function in the vertex function $\Gamma^{(1)}$ (\ref{13}) allows us to take off one of the integrals in the expression for the matrix element (\ref{Matrix}). As the result the usual expression for the matrix element as the overlap integral of the meson wave functions is obtained. The contribution $\Gamma^{(2)}$ (\ref{14}) is significantly more complicated and contains the quasipotential of the quark-antiquark interaction $\mathcal{V}$ (\ref{Ker}) which has nontrivial Lorentz-structure. However, it is possible to use the quasipotential equation (\ref{1}) to perform  one of the integrations in Eq.~(\ref{Matrix}) and thus get again the usual structure of the matrix element as the overlap integral of meson wave functions (for details see Refs. \cite{Ebert2003a}). 

It is convenient to carry out calculations in the rest frame of the decaying hadron, the $D_{(s)}$ meson in the considered case, where the decaying meson momentum ${\bf p}_{D_{(s)}} = 0$. Then the final meson $F$ is moving with the recoil momentum ${\bf\Delta} ={\bf p}_F$ and its wave function should be boosted to the moving reference frame. The wave function of the moving meson $\Psi_{F\,{\bf\Delta} }$ is connected with the wave function in the rest frame $\Psi_{F\,{\bf 0}}$ by the transformation \cite{Ebert2003a}
\begin{equation}
\label{wig}
\Psi_{F\,{\bf\Delta}}({\bf
	p})=D_q^{1/2}(R_{L_{\bf\Delta}}^W)D_{\bar q}^{1/2}(R_{L_{
		\bf\Delta}}^W)\Psi_{F\,{\bf 0}}({\bf p}),
\end{equation}
where $R^W$ is the Wigner rotation, $L_{\bf\Delta}$ is the Lorentz boost
from the meson rest frame to a moving one, and  $D^{1/2}_q(R)$ is 
the rotation matrix  in spinor representation.

\section{DECAY FORM FACTORS}	\label{sec:ff}
	
 The semileptonic $D$ and $D_s$ meson decays to a  pseudoscalar ($P$), vector ($V$), scalar ($S$), axial-vector ($A$) and tensor ($T$) mesons in the standard model are governed by the flavor-changing  current $c\to ql\nu_l$ $(q=s,d)$.
The matrix element $\mathcal{M}$ of this current between meson states  is the product of the leptonic current $L_\mu=\overline{\nu}_l\gamma_\mu(1-\gamma_5)l$ and the matrix element of the hadronic current $H^\mu=\langle{F}| \overline{q}\gamma_\mu(1-\gamma_5)c|{D_{(s)}}\rangle$  	
	\begin{equation}
	\mathcal{M}(D_{(s)}\to Fl\nu_l)=\dfrac{G_F}{\sqrt{2}}V_{cq}H^\mu L_\mu,
	\end{equation}
where  $V_{cq}$ is the corresponding CKM matrix element and $G_F$ is the Fermi constant. The leptonic part is easily
calculated using the lepton spinors and has a simple structure. The hadronic part is significantly more complicated since its calculation requires nonperturbative treatment within QCD. 	

	The hadronic matrix element of the weak current $J^W$ between meson states is parameterized by the following set of the invariant form factors \cite{Ebert2003a,Ebert2010}.

\begin{itemize}	
\item For $D_{(s)}$ transitions to pseudoscalar $P$ mesons	
	\begin{eqnarray}
	\nonumber\langle P(p_F)|\overline{f}\gamma^\mu c|D_{(s)}(p_{D_{(s)}})\rangle &=& f_+(q^2)\left[p^\mu_{D_{(s)}}+p^\mu_P-\dfrac{M_{D_{(s)}}^2-M_P^2}{q^2}q^\mu\right]\\&&+
	f_0(q^2)\dfrac{M_{D_{(s)}}^2-M_P^2}{q^2}q^\mu, \\
	\nonumber \langle P(p_F)|\overline{f}\gamma^\mu \gamma_5c|D_{(s)}(p_{D_{(s)}})\rangle & =& 0,
	\end{eqnarray}
	
\item For $D_{(s)}$ transitions to vector $V$ mesons	
	\begin{eqnarray}
	\nonumber\langle V(p_F)|\overline{f}\gamma^\mu c|D_{(s)}(p_{D_{(s)}})\rangle & =&
		\displaystyle{\frac{2iV(q^2)}{M_{D_{(s)}}+M_V}}{\epsilon}^{\mu \nu \rho \sigma}{\epsilon}^*_\nu p_{D_{(s)}\rho}p_{F\sigma},\\
		\nonumber\langle V(p_F)|\overline{f}\gamma^\mu \gamma_5c|D_{(s)}(p_{D_{(s)}})\rangle &=& 2M_VA_0(q^2)\displaystyle\frac{\epsilon^*q}{q^2}q^\mu
		+(M_{D_{(s)}}+M_V)A_1(q^2)\Big(\epsilon^{*\mu}-\displaystyle\frac{\epsilon^*q}{q^2}q^\mu\Big)\\&&\!\!\!\!\! -A_2(q^2)\displaystyle\frac{\epsilon^*q}{M_{D_{(s)}}+M_V}\left[p^\mu_{D_{(s)}}+p^\mu_V-\dfrac{M^2_{D_{(s)}}-M^2_V}{q^2}q^\mu\right],\qquad
	\end{eqnarray}
\item For $D_{(s)}$ transitions to scalar $S$ mesons		
	\begin{eqnarray}
	\nonumber\langle S(p_F)|\overline{f}\gamma^\mu c|D_{(s)}(p_{D_{(s)}})\rangle &=& 0, \\
	 \langle S(p_F)|\overline{f}\gamma^\mu \gamma_5c|D_{(s)}(p_{D_{(s)}})\rangle &=&f_+(q^2)(p^\mu_{D_{(s)}}+p^\mu_F)+f_-(q^2)(p^\mu_{D_{(s)}}-p^\mu_F),
	\end{eqnarray}
\item For $D_{(s)}$ transitions to axial-vector $A$ mesons		
	\begin{eqnarray}
	\nonumber\langle A(p_F)|\overline{f}\gamma^\mu c|D_{(s)}(p_{D_{(s)}})\rangle &=& (M_{D_{(s)}}+M_A)V_1(q^2){\epsilon}^{*\mu}\\& & +[V_2(q^2)p^\mu_{D_{(s)}}+V_3(q^2)p^\mu_F]\displaystyle \frac{{\epsilon}^*\cdot q}{M_{D_{(s)}}}, \nonumber\\
	\langle A(p_F)|\overline{f}\gamma^\mu \gamma_5c|D_{(s)}(p_{D_{(s)}})\rangle& =& \displaystyle\frac{2iA(q^2)}{M_{D_{(s)}}+M_A}{\epsilon}^{\mu \nu \rho \sigma}{\epsilon}^*_\nu p_{D_{(s)}\rho}p_{F\sigma},
	\end{eqnarray}
\item For $D_{(s)}$ transitions to tensor $T$ mesons		
	\begin{eqnarray}
	\nonumber\langle T(p_F)|\overline{f}\gamma^\mu c|D_{(s)}(p_{D_{(s)}})\rangle &=& \displaystyle\frac{2iV(q^2)}{M_{D_{(s)}}+M_T}{\epsilon}^{\mu \nu \rho \sigma}{\epsilon}^*_{\nu \alpha}\displaystyle\frac{p^\alpha_{D_{(s)}}}{M_{D_{(s)}}}p_{D_{(s)}\rho}p_{F\sigma},\\
	\nonumber\langle T(p_F)|\overline{f}\gamma^\mu \gamma_5c|D_{(s)}(p_{D_{(s)}})\rangle &=& (M_{D_{(s)}}+M_T)A_1(q^2){\epsilon}^{*\mu}_{\alpha}\displaystyle\frac{p^\alpha_{D_{(s)}}}{M_{D_{(s)}}}\\ & &+[A_2(q^2)p^\mu_{D_{(s)}}+A_3(q^2)p^\mu_F]{\epsilon}^*_{\alpha\beta}\displaystyle\frac{p^\alpha_{D_{(s)}}p^\beta_{D_{(s)}}}{M^2_{D_{(s)}}},
	\end{eqnarray}
	\end{itemize}
here $q=	p_{D_{(s)}}-p_F$ and  the following relations among form factors of the $D_{(s)}$ transitions to the radially excited pseudoscalar and vector mesons are satisfied at the maximum recoil~($q^2=0$):
	\[f_+(0)=f_0(0)\quad {\rm and} \quad  A_0(0)=\frac{M_{D_{(s)}}+M_V}{2M_V}A_1(0)-\frac{M_{D_{(s)}}-M_V}{2M_V}A_2(0).\]
	
We use the quasipotential approach and the relativistic quark model discussed in Secs.~\ref{sec:rqm}--\ref{sec:weak} for the calculation of the weak decay matrix elements and transition form factors. We substitute the leading $\Gamma^{(1)}$  (\ref{12}) and subleading $\Gamma^{(2)}$ (\ref{13}) vertex functions in the expression for the matrix element of the weak current between meson states. This matrix element is considered in the rest frame of the decaying $D_{(s)}$ meson, then the boost of the final meson wave function $\Psi_F$ from the rest to moving reference frame with the recoil momentum ${\bf\Delta}=\textbf{p}_F$ should be considered. It is given by Eq.~(\ref{14}). Thus we take into account all relativistic effects including the relativistic contributions of intermediate negative-energy states and relativistic transformations of the meson wave functions. The resulting expressions for the decay form factors have the form of the overlap integrals of initial and final meson wave functions. They are rather cumbersome and are given in Refs. \cite{Ebert2003a,Ebert2010}. For the numerical evaluation of the decay form factors we use the meson wave functions obtained in calculating their mass spectra \cite{Ebert2009,Ebert2010}. This is a significant advantage of our approach since in most of the previous model calculations some phenomenological wave functions (such as Gaussian) were used. Moreover, our relativistic approach allows us to determine the form factor dependence on the transferred momentum squared $q^2$ in the whole accessible kinematical range without additional approximations and extrapolations.
We find that the numerical results for these decay form factors can be approximated with high accuracy by the following expressions.

\begin{itemize}
\item[(a)] For transitions to orbitally excited states
	\begin{equation}
	\label{Ap1}
	F(q^2)=\dfrac{F(0)}{\Big(1-\sigma_1\dfrac{q^2}{M_{D^*_s}^2}+\sigma_2\dfrac{q^4}{M_{D^*_s}^4}+\sigma_3\dfrac{q^6}{M_{D^*_s}^6}\Big)},
	\end{equation}
	where $\sigma_{1,2,3}$ are dimensionless fitted parameters and the mass of the vector $D_s^*$ meson $M_{D^*_s} = 2.112$~GeV was used for normalization.

\item[(b)] For transitions to radially excited states	
\begin{itemize}
\item[(b1)] form factors $f_+(q^2),f_0(q^2),V(q^2),A_0(q^2)$	
	\begin{equation}
	\label{Ap2}
	F(q^2)=\dfrac{F(0)}{\Big(1-\dfrac{q^2}{M^2}\Big)\Big(1-\sigma_1\dfrac{q^2}{M_{D^*_{(s)}}^2}+\sigma_2\dfrac{q^4}{M_{D^*_{(s)}}^4}\Big)},
	\end{equation}
\item[(b2)] form factors $A_1(q^2),A_2(q^2)$
	\begin{equation}
	\label{Ap3}
	F(q^2)=\dfrac{F(0)}{\Big(1-\sigma_1\dfrac{q^2}{M_{D^*_{(s)}}^2}+\sigma_2\dfrac{q^4}{M_{D^*_{(s)}}^4}\Big)},
	\end{equation}
	\end{itemize}
where for the decays governed by the CKM favored $c \to s$ transitions masses of the intermediate $D_s$ mesons are used: $M = M_{D^*_s} = 2.112$~GeV for the form factors $f_+(q^2)$, $f_0(q^2)$, $V(q^2)$ and $M = M_{D_s} = 1.968$~GeV for the form factor $A_0(q^2)$. For decays governed by the CKM suppressed $c \to d$ transitions masses of the intermediate $D$ mesons are employed: $M=M_{D^*}=2.010$~GeV and $M = M_{D} = 1.870$~GeV, respectively. Here $\sigma_{1,2}$ are dimensionless fitted parameters. Note that the same parameterization was previously introduced for the form factors of the semileptonic $D_{(s)}$ decays into the ground state light and strange mesons \cite{Faustov:2022ybm}. 
\end{itemize}

\begin{table}[tb]
	\caption{Form factors of the weak $D$ meson transitions into orbitally excited kaons.}
	\begin{ruledtabular}
		\begin{tabular}{ccccccc}
			\text{Decay}&\text{Form factors}&$F(0)$&$F(q^2_{max})$&$\sigma_1$&$\sigma_2$&$\sigma_3$\\
			\hline
			
			$D\rightarrow K_0^*(1430) l\nu_l$ & $f_+$ & 0.558 & 0.652 & 3.549 & 7.406& -15.86\\
			& $f_-$& -1.370 & -1.418& 0.594 & -2.908&-18.94\\
			$D\rightarrow K_1(1270)l\nu_l$& $A$ &-0.621 &-0.694&1.404&-3.311&61.75\\
			&$V_1$&0.336&0.554&4.242&-1.352&-34.95\\
			&$V_2$&-0.986&-1.111&1.199&3.452&-58.27\\
			&$V_3$&-0.031&-0.052&8.021&53.33&-192.7\\
			$D\rightarrow K_1(1400)l\nu_l$& $A$ &-0.735 &-0.849&2.809&1.514&-5.678\\
			&$V_1$&0.117&0.166&6.461&11.75&-54.04\\
			&$V_2$&-0.648&-0.838&5.268&15.49&-59.91\\
			&$V_3$&2.772&3.084&2.105&0.609&2.295\\
			$D\rightarrow K_2^*(1430)l\nu_l$& $V$ &-1.551 &-1.714&2.158&-0.915&-6.820\\
			&$A_1$&-0.523&-0.536&0.456&-1.535&-8.785\\
			&$A_2$&-0.616&-0.640&0.611&-5.141&-20.83\\
			&$A_3$&-0.005&-0.006&2.559&-2.651&1.689\\
		\end{tabular}\label{FF1}
	\end{ruledtabular}
\end{table}

	The values of form factors $F(0), F(q_{max}^2)$ and parameters $\sigma_{1,2,3}$  fitted to numerically calculated form factors in the whole $q^2$ range are given in Tables~\ref{FF1}--\ref{FF5}. The values of the $q_{max}^2$ were evaluated for the central values of the experimental meson masses given in Tables~\ref{Mass1}--\ref{Mass2}.  We estimate the uncertainties of the calculated form factors which originate from the model parameters and fitting to be less than 5\%.

\begin{table}
	\caption{Form factors of the weak $D$ meson transitions into orbitally excited light mesons.}
	\begin{ruledtabular}
		\begin{tabular}{ccccccc}
			\text{Decay}&\text{Form factor}&$F(0)$&$F(q^2_{max})$&$\sigma_1$&$\sigma_2$&$\sigma_3$\\
			\hline
			$D\rightarrow a_0(1450) l\nu_l$ & $f_+$ & 0.719 & 0.796 & 2.586 & 6.498&-9.296 \\
			& $f_-$& -1.391 & -1.451& 0.769 & -4.681&-17.32\\
			$D\rightarrow a_1(1260)l\nu_l$& $A$ &-0.694 &-1.044&4.309&7.761&-6.376\\
			&$V_1$&0.378&0.374&-0.181&2.057&-28.39\\
			&$V_2$&-1.175&-1.769&3.210&-0.757&-45.39\\
			&$V_3$&2.078&2.845&3.112&0.611&13.76\\
			$D\rightarrow b_1(1235)l\nu_l$& $A$ &0.068 &0.177&14.66&135.5&-536.8\\
			&$V_1$&0.157&0.176&1.210&2.620&-24.51\\
			&$V_2$&-0.611&-0.276&-5.248&-47.69&1469\\
			&$V_3$&-2.697&-3.725&3.125&1.413&-1.162\\
			$D\rightarrow a_2(1320)l\nu_l$& $V$ &-1.602 &-2.117&3.414&-1.628&-10.30\\
			&$A_1$&-0.443&-0.473&0.658&-2.638&-22.09\\
			&$A_2$&-0.680&-0.740&0.009&-10.67&-99.06 \\
			&$A_3$&0.042&0.040&-0.286&4.396&42.58\\
			$D\to f_0(1370)l\nu_l$&$f_+$&0.440&0.541&3.488&8.140&-21.38\\
			&$f_-$&-0.823&-1.129&2.743&-32.14&58.40\\
			$D\to f_0(1500)l\nu_l$&$f_+$&0.349&0.417&6.255&11.20&-83.71\\
			&$f_-$&-0.797&-1.045&7.820&-41.23&282.6\\
			$D\rightarrow f_1(1285)l\nu_l$& $A$ &-0.367 &-0.499&3.648&2.512&4.184\\
			&$V_1$&0.189&0.223&1.281&1.340&-132.4\\
			&$V_2$&0.677&0.770&0.646&0.787&-162.3\\
			&$V_3$&0.974&1.267&3.112&0.611&13.76\\
			$D\rightarrow f_1(1420)l\nu_l$& $A$ &-0.225 &-0.266&3.517&0.235&12.76\\
			&$V_1$&0.083&0.093&2.032&-2.102&-248.1\\
			&$V_2$&-0.342&-0.414&3.990&-0.432&10.41\\
			&$V_3$&0.379&0.425&1.598&-9.888&-226.0\\
			$D\rightarrow h_1(1170)l\nu_l$& $A$ &0.012 &0.164&30.13&343.7&-1330\\
			&$V_1$&0.128&0.134&0.844&0.436&34.88\\
			&$V_2$&-0.539&-0.195&-5.777&-70.91&-1461\\
			&$V_3$&-1.820&-2.611&2.894&1.743&-2.233\\
			$D\rightarrow h_1(1415)l\nu_l$& $A$ &0.014 &0.016&4.566&30.38&200.8\\
			&$V_1$&0.008&0.009&2.367&4.868&30.37\\
			&$V_2$&-0.048&-0.031&-5.280&8.649&2134\\
			&$V_3$&-0.172&-0.213&4.065&1.489&4.157\\
			$D\rightarrow f_2(1270)l\nu_l$& $V$ &-1.063 &-1.375&2.726&-0.751&-11.94\\
			&$A_1$&-0.320&-0.345&0.650&-1.900&-15.94\\
			&$A_2$&-0.472&-0.510&-0.066&-6.635&-76.23\\
			&$A_3$&0.029&0.026&-0.456&3.972&39.25\\
			$D\rightarrow f^\prime_2(1525)l\nu_l$& $V$ &-0.275 &-0.314&4.299&-5.813&-40.57\\
			&$A_1$&-0.044&-0.045&0.589&-10.14&-98.14\\
			&$A_2$&-0.066&-0.073&1.261&-73.62&-236.0\\
			&$A_3$&0.00480&0.00483&0.490&8.853&86.51\\
		\end{tabular}\label{FF3}
	\end{ruledtabular}
\end{table}

\begin{table}[hbt]
	\caption{Form factors of the weak $D_s$ meson transitions into orbitally excited mesons.}
	\begin{ruledtabular}
		\begin{tabular}{ccccccc}
			\text{Decay}&\text{Form factors}&$F(0)$&$F(q^2_{max})$&$\sigma_1$&$\sigma_2$&$\sigma_3$\\
			\hline	
			$D_s\rightarrow K_0^*(1430) l\nu_l$ & $f_+$ & 0.370 & 0.603 &8.647 &55.22&-191.0 \\
			& $f_-$& -1.563 & -1.773&0.933 & -11.65&-20.38\\
			$D_s\rightarrow K_1(1270)l\nu_l$& $A$ &-0.515 &-0.753&2.976&1.982&-0.812\\
			&$V_1$&0.312&0.307&-0.123&0.503&-2.907\\
			&$V_2$&-0.430&-0.959&6.221&22.94&-92.96\\
			&$V_3$&-0.116&-0.139&1.826&3.144&1.149\\
			$D_s\rightarrow K_1(1400)l\nu_l$& $A$ &-0.720 &-0.965&4.174&9.465&-7.611\\
			&$V_1$&0.220&0.198&-1.507&3.180&-32.64\\
			&$V_2$&-1.605&-1.770&1.610&-7.179&159.5\\
			&$V_3$&3.026&3.824&3.129&2.732&4.471\\
			$D_s\rightarrow K_2^*(1430)l\nu_l$& $V$ &-2.351 &-2.710&2.047&1.284&-17.93\\
			&$A_1$&-0.641&-0.689&0.974&-0.866&-11.32\\
			&$A_2$&-0.623&-0.841&2.406&-26.93&38.50\\
			&$A_3$&0.082&0.078&-0.294&3.339&32.79\\
			$D_s\rightarrow f_0(1500) l\nu_l$ & $f_+$ & 0.340 & 0.466 & 7.785& 45.89&-165.1 \\
			& $f_-$& -1.764 & -1.818& 0.587 & -1.426&-10.43\\
			$D_s\rightarrow f_1(1420)l\nu_l$& $A$ &-1.133 &-1.309&2.157&1.769&-2.305\\
			&$V_1$&0.233&0.246&0.587&-3.174&-7.279\\
			&$V_2$&-1.493&-1.881&3.053&-0.158&-20.52\\
			&$V_3$&1.963&2.253&2.059&1.539&-2.042\\
			$D_s\rightarrow h_1(1415)l\nu_l$& $A$ &0.060 &0.075&3.908&17.16&-32.64\\
			&$V_1$&0.120&0.119&0.053&1.780&6.331\\
			&$V_2$&-0.266&-0.383&3.102&-18.38&8.106\\
			&$V_3$&-2.323&-2.461&1.125&1.619&41.53\\
			$D_s\rightarrow f_2(1525)l\nu_l$& $V$ &-2.043 &-2.207&1.608&-0.381&-0.984\\
			&$A_1$&-0.692&-0.716&0.700&-0.683&-3.890\\
			&$A_2$&-0.653&-0.686&0.791&-5.432&-14.75\\
			&$A_3$&0.019&0.018&-0.972&4.896&32.67\\
			$D_s\rightarrow f_0(1370) l\nu_l$ & $f_+$ & 0.114 & 0.199 & 7.784& 45.00&-148.8 \\
			& $f_-$& -0.638 & -0.670& 0.464 & -0.537&-6.317\\
			$D_s\rightarrow f_1(1285)l\nu_l$& $A$ &-0.266 &-0.354&2.723&3.761&-3.848\\
			&$V_1$&0.072&0.071&-0.455&-2.256&-6.420\\
			&$V_2$&-0.366&-0.511&2.675&1.175&-13.20\\
			&$V_3$&0.499&0.621&2.067&2.226&-2.486\\
			$D_s\rightarrow h_1(1170)l\nu_l$& $A$ &0.001 &0.009&20.47&174.9&-519.2\\
			&$V_1$&0.0035&0.0015&-3.109&-106.3&1045\\
			&$V_2$&-0.023&-0.003&-7.318&-96.17&2342\\
			&$V_3$&-0.135&-0.213&2.749&1.880&-3.650\\
			$D_s\rightarrow f_2(1270)l\nu_l$& $V$ &-0.170 &-0.202&1.563&1.087&-2.497\\
			&$A_1$&-0.100&-0.107&0.649&0.254&-1.112\\
			&$A_2$&-0.085&-0.083&-0.262&0.290&-5.157\\
			&$A_3$&0.0033&0.0026&-1.540&4.258&26.66\\
			
		\end{tabular}\label{FF4}
	\end{ruledtabular}
\end{table}

\begin{table}
	\caption{Form factors of the weak $D$ meson transitions into radially excited mesons.}
	\begin{ruledtabular}
		\begin{tabular}{cccccc}
			\text{Decay}&\text{Form factors}&$F(0)$&$F(q^2_{max})$&$\sigma_1$&$\sigma_2$\\
			\hline	
			$D\rightarrow K(1460)l\nu_l$&$f_+$&-0.604&-0.807&2.448&-126.6\\
			&$f_0$&-0.604&-0.394&-10.43&207.4\\
			$D\rightarrow K^*(1680)l\nu_l$&$V$&-1.710&-1.751&2.806&-133.4\\
			&$A_0$&-0.382&-0.385&0.339&32.54\\
			&$A_1$&-0.372&-0.375&0.674&-98.73\\
			&$A_2$&-0.279&-0.275&-1.795&111.2\\
			$D\rightarrow \eta(1295)l\nu_l$&$f_+$&-0.650&-1.036&-0.118&-48.56\\
			&$f_0$&-0.650&-0.449&-3.826&39.15\\
			$D\rightarrow \pi(1300)l\nu_l$&$f_+$&-0.656&-4.296&-0.094&-46.29\\
			&$f_0$&-0.656&-0.228&-3.934&41.48\\
			$D\rightarrow \omega(1420)l\nu_l$&$V$&-1.597&-2.273&1.807&-60.04\\
			&$A_0$&-0.407&-0.469&0.714&-13.94\\
			&$A_1$&-0.425&-0.460&0.150&-25.05\\
			&$A_2$&-0.536&-0.473&-2.275&5.471\\
			$D\rightarrow \rho(1450)l\nu_l$&$V$&-1.771&-2.476&2.470&-104.4\\
			&$A_0$&-0.426&-0.387&0.887&-29.93\\
			&$A_1$&-0.435&-0.442&0.192&-45.12\\
			&$A_2$&-0.500&-0.229&-3.572&99.75\\
			
		\end{tabular}\label{FF2}
	\end{ruledtabular}
\end{table}

\begin{table}[hbt]
	\caption{Form factors of the weak $D_s$ meson transitions into radially excited mesons.}
	\begin{ruledtabular}
		\begin{tabular}{cccccc}
			\text{Decay}&\text{Form factors}&$F(0)$&$F(q^2_{max})$&$\sigma_1$&$\sigma_2$\\
			\hline	
			$D_s\rightarrow K(1460)l\nu_l$&$f_+$&-0.464&-0.977&1.900&-112.5\\
			&$f_0$&-0.464&-0.382&-4.202&12.95\\
			$D_s\rightarrow K^*(1680)l\nu_l$&$V$&-1.281&-1.365&1.546&-93.79\\
			&$A_0$&-0.389&-0.393&0.461&-23.68\\
			&$A_1$&-0.367&-0.372&0.594&-20.55\\
			&$A_2$&-0.208&-0.206&0.730&86.82\\
			$D_s\rightarrow \eta(1475)l\nu_l$&$f_+$&-0.373&-1.035&2.860&-33.36\\
			&$f_0$&-0.373&-0.380&-4.883&-1.898\\
			$D_s\rightarrow \phi(1680)l\nu_l$&$V$&-1.182&-1.601&5.760&-403.7\\
			&$A_0$&-0.357&-0.415&1.515&-200.2\\
			&$A_1$&-0.354&-0.390&2.045&-153.5\\
			&$A_2$&-0.213&-0.162&-10.06&359.6\\
			
		\end{tabular}\label{FF5}
	\end{ruledtabular}
	
	\end{table}

\begin{figure}
	\begin{minipage}[h]{0.47\linewidth}
		\center{\includegraphics[width=1\linewidth]{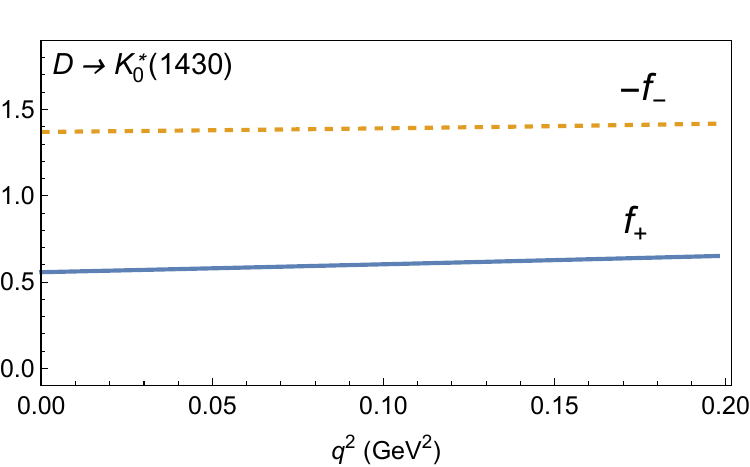}} \\
	\end{minipage}
	\hfill
	\begin{minipage}[h]{0.47\linewidth}
		\center{\includegraphics[width=1\linewidth]{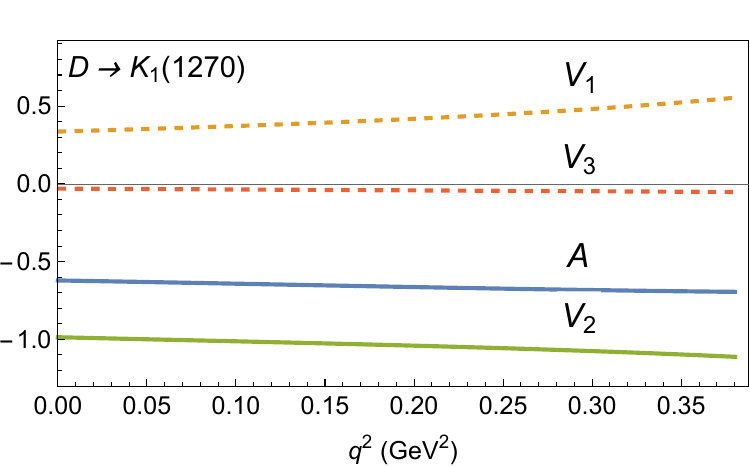}} \\
	\end{minipage}
	\vfill
	\begin{minipage}[h]{0.47\linewidth}
		\center{\includegraphics[width=1\linewidth]{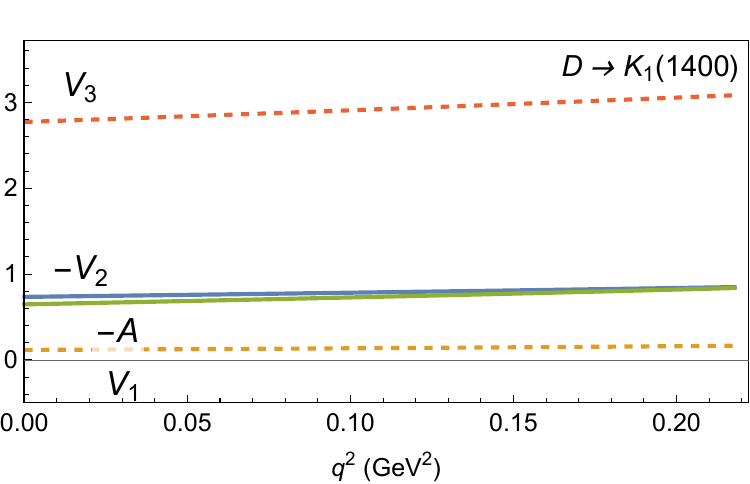}} \\
	\end{minipage}
	\hfill
	\begin{minipage}[h]{0.47\linewidth}
		\center{\includegraphics[width=1\linewidth]{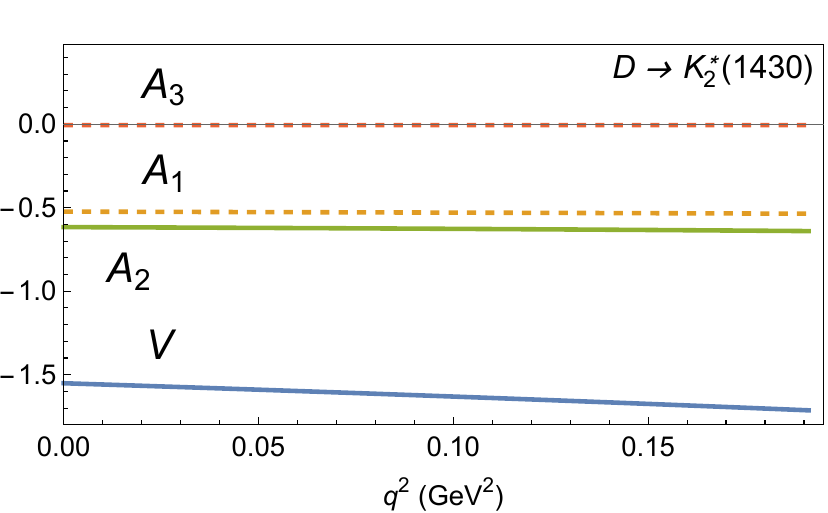}} \\
	\end{minipage}
	\caption{Form factors of the weak transitions of $D$ mesons into orbitally excited kaons}
	\label{GrFF}
\end{figure}

As an example, we plot the form factors of the $D$ meson semileptonic decays into orbitally excited kaons in Fig. \ref{GrFF}. We see from these plots that the kinematical range in these decays is rather small. Indeed, $q^2$ does not exceed 0.4~GeV$^2$ due to the small difference between charm and excited strange meson masses. The same conclusion is also true for other considered decays. It is even smaller for decays into radially excited mesons due to the larger masses of radial excitations. As a result rather small uncertainties in the final mesons masses can significantly change the $q^2$ range. From Tables~\ref{Mass1} and \ref{Mass2} we see that the scalar $0^+$ ($1^3P_0$) and pseudoscalar $0^-$ ($2^1S_0$) meson masses have the largest experimental uncertainties reaching a hundred MeV. The further analysis shows that such uncertainties lead to significant errors in the estimates of the branching fractions.

\section{SEMILEPTONIC DECAYS}\label{sec:sem}

The differential decay rate of the semileptonic $D_{(s)}$ decays can be expressed in the following form~\cite{Ivanov2019}
\begin{eqnarray}\label{Gamma}
\nonumber\dfrac{d\Gamma(D_{(s)}\rightarrow Fl^+\nu_l)}{dq^2d\cos\theta}&=&\dfrac{G^2_F}{(2\pi)^3}|V_{cq}|^2\dfrac{\lambda^{1/2}(q^2-m_l^2)^2}{64M_D^3q^2}
\Big[(1+\cos^2\theta)\mathcal{H}_U +2\sin^2\theta \mathcal{H}_L +2\cos\theta \mathcal{H}_P\\
&&+2\delta_l\Big(\sin^2\theta \mathcal{H}_U+2\cos^2\theta \mathcal{H}_L+2\mathcal{H}_S-4\cos\theta \mathcal{H}_{SL}\Big) \Big],
\end{eqnarray}	
where $\lambda\equiv\lambda(M_{D_{(s)}}^2,M_F^2,q^2)= M_{D_{(s)}}^4+M_F^4+q^4-2(M_{D_{(s)}}^2M_F^2+M_{D_{(s)}}^2q^2+M^2_Fq^2)$, $m_l$ is
the lepton mass, $\delta_l=\frac{m_l^2}{2q^2}$, and the polar angle $\theta$ is the angle between the momentum of the charged lepton in the rest frame of the intermediate $W$-boson and the direction opposite to the final $F$ meson momentum in the rest frame of $D_{(s)}$. The bilinear combinations $\mathcal{H}_I$ ($I=U,L,P,S,SL$) of the helicity components of the hadronic tensor are defined by
\begin{eqnarray}\label{hc}
&&\mathcal{H}_U=|H_+|^2+|H_-|^2,\qquad \mathcal{H}_L=|H_0|^2,\qquad \mathcal{H}_P=|H_+|^2-|H_-|^2,\cr
&& \mathcal{H}_S=|H_t|^2,\qquad\qquad \qquad\mathcal{H}_{SL}=\Re(H_0H^\dagger_t),
\end{eqnarray}
and the helicity amplitudes are expressed through invariant form factors~\cite{Ivanov2019,Ebert2010,Faustov:2019mqr}.
\begin{itemize}

\item For $D_{(s)}\to P$ transitions
\begin{eqnarray} \label{dp}
H_\pm =0, \qquad H_0=\dfrac{\lambda^{1/2}}{\sqrt{q^2}}f_+(q^2),\qquad H_t=\dfrac{1}{\sqrt{q^2}}\Big(M^2_{D_{(s)}}-M^2_P\Big)f_0(q^2).
\end{eqnarray}

\item For $D_{(s)}\to V$ transitions
\begin{eqnarray}\label{dv}
H_\pm&=&\dfrac{\lambda^{1/2}}{M_{D_{(s)}}+M_V}\left[V(q^2)\mp \dfrac{(M_{D_{(s)}}+M_V)^2}{\lambda^{1/2}}A_1(q^2)\right],\cr
H_0&=&\dfrac{1}{2M_V\sqrt{q^2}}\left[(M_{D_{(s)}}+M_V)(M^2_{D_{(s)}}-M_V^2-q^2)A_1(q^2)-\dfrac{\lambda}{M_{D_{(s)}}+M_V}A_2(q^2)\right],\cr
H_t&=&\dfrac{\lambda^{1/2}}{\sqrt{q^2}}A_0(q^2).
\end{eqnarray}

\item For $D_{(s)}\to S$ transitions
\begin{eqnarray}\label{ds}
H_\pm&=&0,\qquad H_0=\dfrac{\lambda^{1/2}}{\sqrt{q^2}}f_+(q^2),\cr H_t&=&\dfrac{1}{\sqrt{q^2}}[(M^2_{D_{(s)}}-M^2_S)f_+(q^2)+q^2f_-(q^2)].
\end{eqnarray}

\item For $D_{(s)}\to A$ transitions
\begin{eqnarray}\label{da}
\nonumber H_\pm&=&(M_{D_{(s)}}+M_{A})V_1(q^2)\pm\dfrac{\lambda^{1/2}}{M_{D_{(s)}}+M_{A}}A(q^2),\\
\nonumber H_0&=&\dfrac{1}{2M_{A}\sqrt{q^2}}\Big\{(M_{D_{(s)}}+M_{A})(M^2_{D_{(s)}}-M^2_{A}-q^2)V_1(q^2)\\
\nonumber &&\qquad\qquad\quad+\dfrac{\lambda}{2M_{D_{(s)}}}[V_2(q^2)+V_3(q^2)] \Big\},\\
\nonumber H_t&=&\dfrac{\lambda^{1/2}}{2M_{A}\sqrt{q^2}}\Bigg\{(M_{D_{(s)}}+M_{A})V_1(q^2)+\dfrac{M_{D_{(s)}}^2-M^2_{A}}{2M_{D_{(s)}}}[V_2(q^2)+V_3(q^2)]\\&&\qquad\qquad\quad+\dfrac{q^2}{2M_{D_{(s)}}}[V_2(q^2)-V_3(q^2)] \Bigg\}.
\end{eqnarray}

\item For $D_{(s)}\to T$ transitions
\begin{eqnarray}\label{dt}
\nonumber H_\pm&=&\dfrac{\lambda^{1/2}}{2\sqrt{2}M_{D_{(s)}}M_T}\left\{(M_{D_{(s)}}+M_T)A_1(q^2)\pm\dfrac{\lambda^{1/2}}{M_{D_{(s)}}+M_T}V(q^2) \right\},\\
\nonumber H_0&=&\dfrac{\lambda^{1/2}}{2\sqrt{6}M_{D_{(s)}}M_{T}^2\sqrt{q^2}}\Big\{(M_{D_{(s)}}+M_{T})(M^2_{D_{(s)}}-M^2_{T}-q^2)A_1(q^2)\\
&&\nonumber\qquad\qquad\qquad\qquad\ \ +\dfrac{\lambda}{2M_{D_{(s)}}}[A_2(q^2)+A_3(q^2)] \Big\},\\
\nonumber H_t&=&\sqrt{\dfrac{2}{3}}\dfrac{\lambda}{4M_DM_{T}^2\sqrt{q^2}}\Bigg\{(M_{D_{(s)}}+M_T)A_1(q^2)\\
&&\qquad\qquad+\dfrac{M_{D_{(s)}}^2-M^2_T}{2M_{D_{(s)}}}[A_2(q^2)+A_3(q^2)]+\frac{q^2}{2M_{D_{(s)}}}[A_2(q^2)-A_3(q^2)] \Bigg\}.\quad
\end{eqnarray}
\end{itemize}
Here the subscripts $\pm,0,t$ denote transverse, longitudinal, and time helicity components, respectively.

The expression for the differential decay rate (\ref{Gamma}) normalized by the decay rate  integrated over $\cos\theta$,
\begin{equation}
  \label{eq:dg}
d\Gamma/dq^2\equiv  \frac{d\Gamma(D_{(s)}\to
  F l^+\nu_l)}{dq^2}=\frac{G_F^2}{(2\pi)^3}
  |V_{cq}|^2\frac{\lambda^{1/2}q^2}{24M_{{D_{(s)}}}^3}\left(1-\frac{m_\ell^2}{q^2}\right)^2\mathcal{H}_{\rm tot},
\end{equation}
 can be rewritten as
\begin{eqnarray}
\dfrac{1}{d\Gamma/dq^2}\dfrac{d\Gamma(D_{(s)}\to Fl^+\nu_l)}{dq^2d(\cos\theta)}=\dfrac{1}{2}\Big[1-\frac{1}{3}C^l_F(q^2)\Big]+A_{FB}(q^2)\cos\theta+\frac{1}{2}C^l_F(q^2)\cos^2\theta,
\end{eqnarray}
where the total helicity structure
\begin{eqnarray}
\mathcal{H}_{\rm tot}=\mathcal{H}_U+\mathcal{H}_L+\delta_l(\mathcal{H}_U+\mathcal{H}_L+3\mathcal{H}_S).
\end{eqnarray}
The forward-backward asymmetry is defined by
\begin{eqnarray}
\label{Afb}
A_{FB}(q^2)=\frac{\int_{0}^{1} d\cos\theta \frac{d\Gamma}{dq^2d\cos\theta}-\int_{-1}^{0} d\cos\theta \frac{d\Gamma}{dq^2d\cos\theta}}{\int_{0}^{1} d\cos\theta \frac{d\Gamma}{dq^2d\cos\theta}+\int_{-1}^{0} d\cos\theta \frac{d\Gamma}{dq^2d\cos\theta}}=\frac{3}{4}\frac{\mathcal{H}_P-4\delta_l\mathcal{H}_{SL}}{\mathcal{H}_{\rm tot}},
\end{eqnarray}
and the lepton-side convexity parameter, which is the second derivative of the distribution  over $\cos\theta$, is given by
\begin{equation}
\label{Clf}
C^l_F(q^2)=\frac{3}{4}(1-2\delta_l)\frac{\mathcal{H}_U-2\mathcal{H}_L}{\mathcal{H}_{\rm tot}}.
\end{equation}

Other useful observables are the longitudinal polarization of the final charged lepton $l$ defined by 
\begin{eqnarray}
\label{Pll}
P^l_L(q^2)=\frac{\mathcal{H}_U+\mathcal{H}_L-\delta_l(\mathcal{H}_U+\mathcal{H}_L+3\mathcal{H}_S)}{\mathcal{H}_{\rm tot}}.
\end{eqnarray}
and its transverse polarization
\begin{eqnarray}
\label{Plt}
P^l_T(q^2)=-\dfrac{3\pi\sqrt{\delta_l}}{4\sqrt{2}}\frac{\mathcal{H}_P+2\mathcal{H}_{SL}}{\mathcal{H}_{\rm tot}}.
\end{eqnarray}

For the semileptonic $D_{(s)}$ decays to the vector $V$ meson, which then decays to two pseudoscalar mesons $V\to P_1P_2$, the differential distribution in the angle $\theta^*$, defined as the polar angle between the vector meson $V$ momentum in the $D_{(s)}$ meson rest frame and the pseudoscalar meson $P_1$ momentum in the rest frame of the vector meson $V$, is given by \cite{Ivanov2019}
\begin{equation}
  \label{eq:mpdr}
\frac1{d\Gamma/dq^2} \frac{d\Gamma(D_{(s)}\to
  V(\to P_1P_2)l^+\nu_l)}{dq^2d(\cos\theta^*)} =
\frac34\left[2F_L(q^2)\cos^2\theta^* +F_T(q^2)\sin^2\theta^*\right].
\end{equation}
Here the longitudinal polarization fraction of the final vector meson has
the form
\begin{equation}
\label{Fl}
F_L(q^2)=\frac{\mathcal{H}_L+\delta_l(\mathcal{H}_L+3\mathcal{H}_S)}{\mathcal{H}_{\rm tot}},
\end{equation}
and its transverse polarization fraction $F_T(q^2)=1- F_L(q^2)$.

\section{RESULTS AND DISCUSSION}\label{sec:results}

\begin{figure}
	\begin{minipage}[h]{0.47\linewidth}
		\center{\includegraphics[width=1\linewidth]{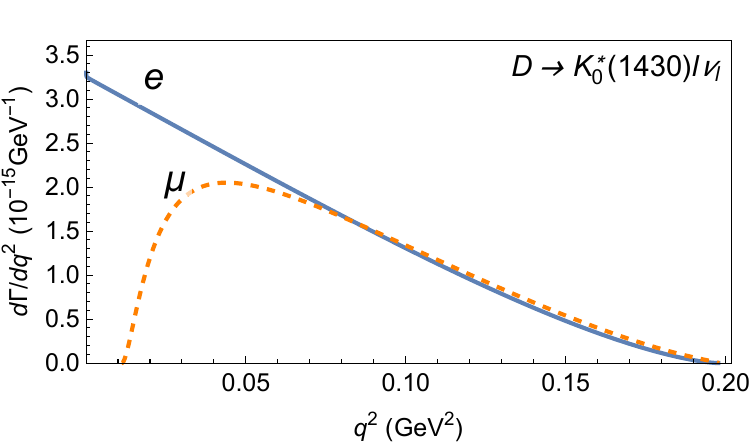}} \\
	\end{minipage}
	\hfill
	\begin{minipage}[h]{0.47\linewidth}
		\center{\includegraphics[width=1\linewidth]{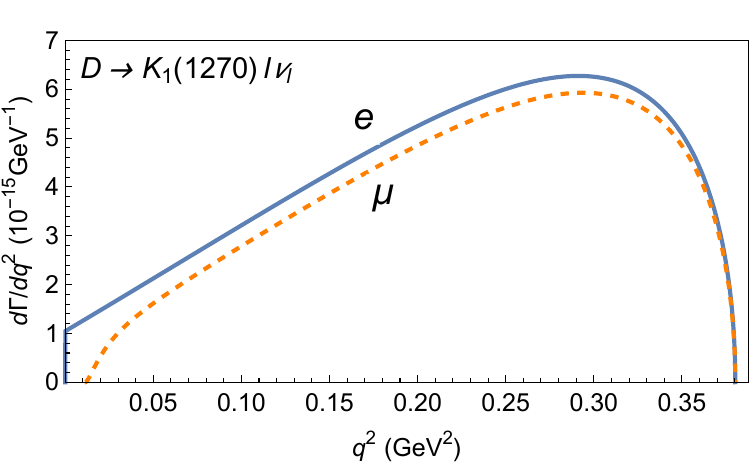}} \\
	\end{minipage}
	\vfill
	\begin{minipage}[h]{0.47\linewidth}
		\center{\includegraphics[width=1\linewidth]{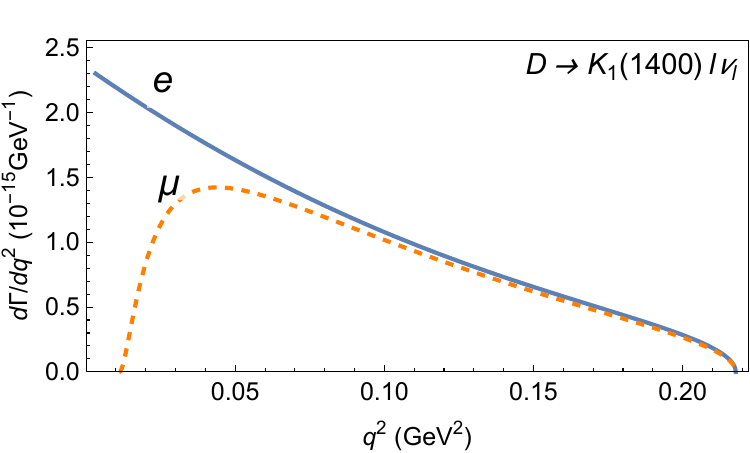}} \\
	\end{minipage}
	\hfill
	\begin{minipage}[h]{0.47\linewidth}
		\center{\includegraphics[width=1\linewidth]{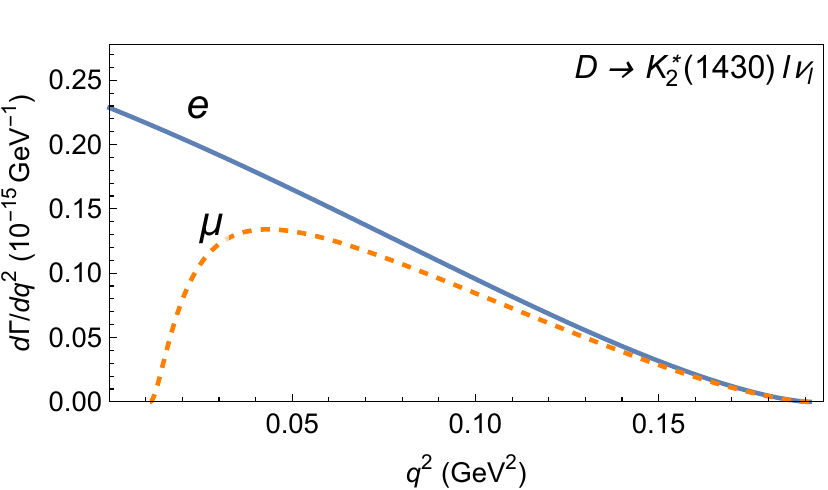}} \\
	\end{minipage}
	\caption{Differential decay rates  of the semileptonic $D$ decays into orbitally excited kaons.}
	\label{GrGamma}
\end{figure}

We substitute the form factors calculated in Sec.~\ref{sec:ff} in the expressions for the helicity amplitudes (\ref{dp})--(\ref{dt}). Then using helicity components of the hadronic tensor (\ref{hc}) and expression (\ref{Gamma}) for the differential decay rate we  evaluate the branching fractions of the semileptonic $D_{(s)}$ decays into excited strange and light mesons. The obtained predictions  are given in Tables~\ref{Br1}--\ref{Br4}. As it is expected, the largest branching fractions are obtained for the CKM favored $c\to s$ transitions, which reach an order of about 2.5\%  for the $D\to K_1(1270)l\nu_l$ decay. In general, such transitions have larger branching fractions for the $D$ then for $D_s$ meson decays due to the lighter mass of the spectator quark and broader $q^2$ range. In Fig.~\ref{GrGamma} we plot the differential decay rates of the semileptonic $D$ decays to orbitally excited kaons. Plots both for decays with the positron (solid line) and muon (dashed line) are presented.

\begin{table}
	\caption{Branching fractions of the semileptonic $D$ decays into orbitally excited mesons ($\times 10^{-5}$).}
	\begin{ruledtabular}
		\begin{tabular}{cc|cc}		
			\text{Decay}&\text{Br}&\text{Decay}&\text{Br}\\
			\hline
			$D^+\rightarrow \overline{K}_0^*(1430)^0 e^+\nu_e$&$44^{+29}_{-19}$&$D^+\rightarrow a_0(1450)^0 e^+\nu_e$ &$1.57^{+0.69}_{-0.53}$\\
			$D^0\rightarrow K_0^*(1430)^- e^+\nu_e$&$17^{+11}_{-8}$&$D^0\rightarrow a_0(1450)^- e^+\nu_e$ &$1.18^{+0.51}_{-0.39}$\\
			$D^+\rightarrow \overline{K}_0^*(1430)^0 \mu^+\nu_\mu$&$32^{+25}_{-15}$&$D^+\rightarrow a_0(1450)^0 \mu^+\nu_\mu$ &$1.13^{+0.58}_{-0.42}$\\
			$D^0\rightarrow K_0^*(1430)^- \mu^+\nu_\mu$&$12^{+9}_{-6}$&	$D^0\rightarrow a_0(1450)^- \mu^+\nu_\mu$ &$0.85^{+0.43}_{-0.32}$\\
			$D^+\rightarrow \overline{K}_1(1270)^0e^+\nu_e$&$257\pm28$&$D^+\rightarrow a_1(1260)^0e^+\nu_e$&$12.4^{+4.7}_{-3.6}$\\
			$D^0\rightarrow K_1(1270)^-e^+\nu_e$&$97\pm 11$&$D^0\rightarrow a_1(1260)^-e^+\nu_e$&$9.5^{+3.5}_{-2.7}$\\
			$D^+\rightarrow \overline{K}_1(1270)^0\mu^+\nu_\mu$&$231\pm25$&$D^+\rightarrow a_1(1260)^0\mu^+\nu_\mu$&$10.5^{+4.2}_{-3.2}$\\
			$D^0\rightarrow K_1(1270)^-\mu^+\nu_\mu$&$87\pm10$&$D^0\rightarrow a_1(1260)^-\mu^+\nu_\mu$&$8.0^{+3.1}_{-2.5}$\\
			$D^+\rightarrow \overline{K}_1(1400)^0e^+\nu_e$&$37\pm4$&	$D^+\rightarrow b_1(1235)^0e^+\nu_e$&$3.39\pm 0.40$\\
			$D^0\rightarrow K_1(1400)^-e^+\nu_e$&$13.8\pm 1.5$&	$D^0\rightarrow b_1(1235)^-e^+\nu_e$&$2.56\pm0.30$\\
			$D^+\rightarrow \overline{K}_1(1400)^0\mu^+\nu_\mu$&$27\pm3$&$D^+\rightarrow b_1(1235)^0\mu^+\nu_\mu$&$2.76\pm0.32$\\
			$D^0\rightarrow K_1(1400)^-\mu^+\nu_\mu$&$9.9\pm1.1$&	$D^0\rightarrow b_1(1235)^-\mu^+\nu_\mu$&$2.08\pm0.24$\\
			$D^+\rightarrow \overline{K}_2^*(1430)^0e^+\nu_e$&$3.15\pm0.35$&$D^+\rightarrow a_2(1320)^0e^+\nu_e$&$0.345\pm0.036$\\
			$D^0\rightarrow K_2^*(1430)^-e^+\nu_e$&$1.26\pm0.14$&$D^0\rightarrow a_2(1320)^-e^+\nu_e$&$0.260\pm0.028$\\
			$D^+\rightarrow \overline{K}_2^*(1430)^0\mu^+\nu_\mu$&$2.02\pm0.22$&	$D^+\rightarrow a_2(1320)^0\mu^+\nu_\mu$&$0.258\pm0.027$\\
			$D^0\rightarrow K_2^*(1430)^-\mu^+\nu_\mu$&$0.81\pm0.09$&$D^0\rightarrow a_2(1320)^-\mu^+\nu_\mu$&$0.194\pm0.020$\\
			$D^+\to f_0(1370)^0e^+\nu_e$&$\{0.6,9.7\}$&$D^+\to f_0(1500)^0e^+\nu_e$&$0.29^{+0.12}_{-0.10}$\\
			$D^+\to f_0(1370)^0\mu^+\nu_\mu$&$\{0.4,8.5\}$&$D^+\to f_0(1500)^0\mu^+\nu_\mu$&$0.18^{+0.10}_{-0.07}$\\
			$D^+\to f_1(1285)^0e^+\nu_e$&$8.7\pm1.0$&$D^+\to f_1(1420)^0e^+\nu_e$&$0.169\pm0.017$\\
			$D^+\to f_1(1285)^0\mu^+\nu_\mu$&$7.1\pm0.8$&$D^+\to f_1(1420)^0\mu^+\nu_\mu$&$0.126\pm0.013$\\
			$D^+\to f_2(1270)^0e^+\nu_e$&$0.63\pm0.07$&$D^+\to f^{'}_2(1525)^0e^+\nu_e$&$2.56\pm0.28(10^{-4})$\\
			$D^+\to f_2(1270)^0\mu^+\nu_\mu$&$0.49\pm0.05$&$D^+\to f^{'}_2(1525)^0\mu^+\nu_\mu$&$1.32\pm0.15(10^{-4})$\\
			$D^+\to h_1(1170)^0e^+\nu_e$&$6.4\pm0.7$&$D^+\to h_1(1415)^0e^+\nu_e$&$2.9\pm0.4(10^{-3})$\\
			$D^+\to h_1(1170)^0\mu^+\nu_\mu$&$5.4\pm0.6$&$D^+\to h_1(1415)^0\mu^+\nu_\mu$&$2.05\pm0.27(10^{-3})$\\
		\end{tabular}\label{Br1}
	\end{ruledtabular}
\end{table}

\begin{table}[hbt]
	\caption{Branching fractions of the semileptonic $D$ decays into radially excited mesons ($\times 10^{-5}$).}
	\begin{ruledtabular}
		\begin{tabular}{cc|cc}
			\text{Decay}&\text{Br}&\text{Decay}&\text{Br}\\
			\hline
			$D^+\rightarrow \overline{K}(1460)^0e^+\nu_e$&$27.7^{+7.5}_{-6.4}$&$D^+\rightarrow \pi(1300)^0e^+\nu_e$&$5.0^{+6.4}_{-3.0}$\\
			$D^0\rightarrow K(1460)^-e^+\nu_e$&$10.3^{+2.9}_{-2.4}$&$D^0\rightarrow \pi(1300)^-e^+\nu_e$&$3.8^{+4.9}_{-2.3}$\\
			$D^+\rightarrow \overline{K}(1460)^0\mu^+\nu_\mu$&$17.9^{+5.9}_{-4.7}$&$D^+\rightarrow \pi(1300)^0\mu^+\nu_\mu$&$4.1^{+5.8}_{-2.6}$\\
			$D^0\rightarrow K(1460)^-\mu^+\nu_\mu$&$6.6^{+2.1}_{-1.7}$&$D^0\rightarrow \pi(1300)^-\mu^+\nu_\mu$&$3.1^{+4.4}_{-2.0}$\\
			$D^+\rightarrow \overline{K}^*(1680)^0e^+\nu_e$&$0.33^{+0.24}_{-0.15}$&$D^+\rightarrow \rho(1450)^0e^+\nu_e$&$1.35^{+0.47}_{-0.37}$\\
			$D^0\rightarrow K^*(1680)^-e^+\nu_e$&$0.11^{+0.10}_{-0.06}$&$D^0\rightarrow \rho(1450)^-e^+\nu_e$&$1.01^{+0.36}_{-0.28}$\\
			$D^+\rightarrow \overline{K}^*(1680)^0\mu^+\nu_\mu$&$0.022^{+0.053}_{-0.018}$&$D^+\rightarrow \rho(1450)^0\mu^+\nu_\mu$&$0.99^{+0.39}_{-0.30}$\\
			$D^0\rightarrow K^*(1680)^-\mu^+\nu_\mu$&$0.006^{+0.017}_{-0.005}$&$D^0\rightarrow \rho(1450)^-\mu^+\nu_\mu$&$0.73^{+0.30}_{-0.22}$\\
			$D^+\rightarrow \eta(1295)^0e^+\nu_e$&$5.14\pm0.46$&$D^+\rightarrow \omega(1420)^0e^+\nu_e$&$2.3^{+2.0}_{-1.2}$\\
			$D^+\rightarrow \eta(1295)^0\mu^+\nu_\mu$&$4.24\pm0.38$&$D^+\rightarrow \omega(1420)^0\mu^+\nu_\mu$&$1.8^{+1.8}_{-1.1}$\\
		\end{tabular}\label{Br2}
	\end{ruledtabular}
\end{table}

\begin{table}[hbt]
	\caption{Branching fractions of semileptonic $D_s$ decays into orbitally excited mesons($\times 10^{-5}$).}
	\begin{ruledtabular}
		\begin{tabular}{cc|cc}
			\text{Decay}&\text{Br}&\text{Decay}&\text{Br}\\
			\hline
			$D_s^+\rightarrow K_0^*(1430)^0e^+\nu_e$&$1.55^{+0.90}_{-0.61}$&$D_s^+\rightarrow f_0(1500)^0e^+\nu_e$&$9.2^{+2.8}_{-2.3}$\\
			$D_s^+\rightarrow K_0^*(1430)^0\mu^+\nu_\mu$&$1.27^{+0.81}_{-0.53}$&$D_s^+\rightarrow f_0(1500)^0\mu^+\nu_\mu$&$6.6^{+2.4}_{-1.8}$\\
			$D_s^+\rightarrow K_1(1270)^0e^+\nu_e$&$7.2\pm0.8$&$D_s^+\rightarrow f_1(1420)^0e^+\nu_e$&$39.9\pm4.0$\\
			$D_s^+\rightarrow K_1(1270)^0\mu^+\nu_\mu$&$6.5\pm0.7$&$D_s^+\rightarrow f_1(1420)^0\mu^+\nu_\mu$&$32.1\pm3.3$\\
			$D_s^+\rightarrow K_1(1400)^0e^+\nu_e$&$3.4\pm0.4$&$D_s^+\rightarrow h_1(1415)^0e^+\nu_e$&$12.6\pm1.9$\\
			$D_s^+\rightarrow K_1(1400)^0\mu^+\nu_\mu$&$2.7\pm0.3$&$D_s^+\rightarrow h_1(1415)^0\mu^+\nu_\mu$&$9.8\pm1.5$\\
			$D_s^+\rightarrow K_2^*(1430)^0e^+\nu_e$&$0.47\pm0.05$&$D_s^+\rightarrow f_2'(1525)^0e^+\nu_e$&$2.90\pm0.32$\\
			$D_s^+\rightarrow K_2^*(1430)^0\mu^+\nu_\mu$&$0.35\pm0.04$&$D_s^+\rightarrow f_2'(1525)^0\mu^+\nu_\mu$&$1.92\pm0.21$\\
			$D_s^+\to f_0(1370)^0e^+\nu_e$&$\{1.30,16.8\}$&$D_s^+\to f_1(1285)^0e^+\nu_e$&$11.7\pm1.8$\\
			$D_s^+\to f_0(1370)^0\mu^+\nu_\mu$&$\{1,15\}$&$D_s^+\to f_1(1285)^0\mu^+\nu_\mu$&$10.1\pm1.2$\\
			$D_s^+\to h_1(1170)^0e^+\nu_e$&$0.98\pm0.11$&$D_s^+\to f_2(1270)^0e^+\nu_e$&$1.43\pm0.15$\\
			$D_s^+\to h_1(1170)^0\mu^+\nu_\mu$&$0.85\pm0.10$&$D_s^+\to f_2(1270)^0\mu^+\nu_\mu$&$1.18\pm0.12$\\
		\end{tabular}\label{Br3}
	\end{ruledtabular}
\end{table}
\begin{table}[hbt]
	\caption{Branching fractions of the semileptonic $D_s$ decays into radially excited mesons. ($\times 10^{-5}$).}
	\begin{ruledtabular}
		\begin{tabular}{cc|cc}
			\text{Decay}&\text{Br}&\text{Decay}&\text{Br}\\
			\hline
			$D_s^+\rightarrow K(1460)^0e^+\nu_e$&$1.35^{+0.34}_{-0.27}$&$D_s^+\rightarrow \eta(1475)^0e^+\nu_e$&$16.6\pm1.9$\\
			$D_s^+\rightarrow K(1460)^0\mu^+\nu_\mu$&$1.06^{+0.29}_{-0.23}$&$D_s^+\rightarrow \eta(1475)^0\mu^+\nu_\mu$&$12.9\pm1.5$\\
			$D_s^+\rightarrow K^*(1680)^0e^+\nu_e$&$0.093^{+0.037}_{-0.028}$&$D_s^+\rightarrow \phi(1680)^0e^+\nu_e$&$3.5^{+1.5}_{-1.1}$\\
			$D_s^+\rightarrow K^*(1680)^0\mu^+\nu_\mu$&$0.039^{+0.023}_{-0.015}$&$D_s^+\rightarrow \phi(1680)^0\mu^+\nu_\mu$&$1.9^{+1.0}_{-0.8}$\\
		\end{tabular}\label{Br4}
	\end{ruledtabular}
\end{table}

The  uncertainty of the presented predictions originate from  the experimental errorbars in the final meson masses, in the elements of the CKM-matrix and meson lifetimes, as well as from the uncertainties in the form factor calculations (see Sec.~\ref{sec:ff}). From Tables~\ref{Br1}, \ref{Br3} we see that predictions for the semileptonic decays to the scalar mesons have largest uncertainties which for some decays exceed 50\%. These uncertainties  mostly originate from the poorly measured masses of scalar mesons, which in most cases are the heaviest masses of the orbitally excited ($P$-wave) mesons (see Tables~\ref{Mass1} and \ref{Mass2}). As a result the accessible $q^2$ range has the largest relative errors. For example, for charmed mesons decays into the $f_0(1370)$ meson only intervals of the branching fractions can be given. The uncertainties of the branching fractions of the semileptonic decays into radially excited mesons given in Tables~\ref{Br2}, \ref{Br4} are also large.  They are especially large for decays to the $\pi(1300)$ and vector mesons  due to the poorly measured mass of the $\pi(1300)$ and narrowest $q^2$ range for decays into vector mesons.

As already noted, relativistic effects are very important in our calculations. To clarify their role we calculated, as an example, the form factors of the $D\to K_1(1270)$ transition neglecting all relativistic contributions.  We then used these form factors to evaluate the corresponding branching fractions in the nonrelativistic limit. The results are the following: for the charged $D^+$ meson decays $Br(D^{+}\to \overline{K}_1(1270)^0e^+\nu_e)\approx28\times10^{-5}$; $Br(D^{+}\to \overline{K}_1(1270)^0\mu^+\nu_\mu)\approx22\times10^{-5}$ and for the neutral $D^0$ meson decays $Br(D^0\to K_1(1270)^-e^+\nu_e)\approx11\times10^{-5}$;  $Br(D^0\to K_1(1270)^-\mu^+\nu_\mu)\approx9\times10^{-5}$. Comparing these values with the relativistic predictions and experimental data \cite{ParticleDataGroup:2024cfk} given in Table~\ref{BrCom1}, we see that nonrelativistic calculations underestimate the decay branching fractions by almost an order of magnitude.

\begin{table}[hbt]
	\caption{Comparison of the calculated branching fractions with other theoretical predictions and experimental data for the semileptonic charm meson decays into axial vector and tensor strange mesons ($\times 10^{-5}$).}
	\begin{ruledtabular}
		\begin{tabular}{@{}ccccc@{}c@{\!\!\!}c@{}}
			\text{Decay}&\text{Our}&\cite{Qiao2024}&\cite{Khosravi2009}&\cite{Momeni2019}&\cite{Cheng2017}&\text{PDG}\cite{ParticleDataGroup:2024cfk}, BESIII\cite{BESIII:2025yot}\\
			\hline
			
			$D^+\rightarrow \overline{K}_1(1270)^0e^+\nu_e$ & $257\pm28$ &$243\pm27$&$270\pm25$&$1686\pm27$&$320\pm40$&$230\pm26\pm25^{+18}_{-21}$ \\
			$D^+\rightarrow \overline{K}_1(1270)^0\mu^+\nu_\mu$&$231\pm25$&$210\pm23$&&&$260\pm30$&$236\pm20\pm48^{+18}_{-27}$\\
			$D^0\rightarrow K_1(1270)^-e^+\nu_e$&$97\pm11$&$93\pm10$&$103\pm10$&$678\pm12$&&$101\pm18$\\
			$D^0\rightarrow K_1(1270)^-\mu^+\nu_\mu$&$87\pm10$&$81\pm9$&&&&$78\pm11\pm15^{+5}_{-9}$\\
			$D_s^+\rightarrow K_1(1270)^0e^+\nu_e$&$7.2\pm0.8$&$11.77\pm1.39$&$20.9\pm2.4$&$166\pm5$&$17\pm2$&$<41$\\
			$D_s^+\rightarrow K_1(1270)^0\mu^+\nu_\mu$&$6.5\pm0.7$&$10.57\pm1.25$&&&$15\pm2$&\\
			$D^+\rightarrow \overline{K}_1(1400)^0e^+\nu_e$&$37\pm4$&$5\pm5$&$475\pm29$&$128\pm8$&$ \{0.5;2.0\} $&\\
			$D^+\rightarrow \overline{K}_1(1400)^0\mu^+\nu_\mu$&$27\pm3$&$4\pm4$&&&$\{0.4;1.7\}$&\\
			$D^0\rightarrow K_1(1400)^-e^+\nu_e$&$13.8\pm1.5$&$2\pm2$&$178\pm15$&$82\pm5$&&\\
			$D^0\rightarrow K_1(1400)^-\mu^+\nu_\mu$&$9.9\pm1.1$&$2\pm2$&&&&\\	
			$D_s^+\rightarrow K_1(1400)^0e^+\nu_e$&$3.4\pm0.4$&$0.32\pm0.32$&$58.8\pm3.4$&$16\pm2$&$\{0.05;0.14\}$&\\
			$D_s^+\rightarrow K_1(1400)^0\mu^+\nu_\mu$&$2.7\pm0.3$&$0.27\pm0.27$&&&$\{0.05;0.12\}$&\\			
			$D^+\rightarrow \overline{K}_2^*(1430)^0e^+\nu_e$&$3.15\pm0.35$&$3.40\pm1.21$&&&&\\		
			$D^+\rightarrow \overline{K}_2^*(1430)^0\mu^+\nu_\mu$&$2.02\pm0.22$&$2.25\pm0.76$&&&&\\
			$D^0\rightarrow K_2^*(1430)^-e^+\nu_e$&$1.26\pm0.14$&$1.36\pm0.49$&&&&\\
			$D^0\rightarrow K_2^*(1430)^-\mu^+\nu_\mu$&$0.81\pm0.09$&$0.91\pm0.31$&&&&\\
			$D^+_s\to K^*_2(1430)^0e^+\nu_e$&$0.47\pm0.05$&$0.34\pm0.13$&&&&\\
			$D^+_s\to K^*_2(1430)^0\mu^+\nu_\mu$&$0.35\pm0.04$&$0.26\pm0.09$&&&&\\	
		\end{tabular}\label{BrCom1}
	\end{ruledtabular}
\end{table}
\begin{table}[hbt]
	\caption{Comparison of the calculated branching fractions with other theoretical predictions and experimental data for $D_{(s)}\to K^*_0(1430)l\nu_l$ ($\times 10^{-5}$).}
	\begin{ruledtabular}
		\begin{tabular}{ccccccc}
			\text{Decay}&\text{Our}&\cite{Yang2006}&S1\cite{Yang2024}&S2\cite{Yang2024}&\cite{Huang2023}&\text{PDG}\cite{ParticleDataGroup:2024cfk}\\
			\hline
			$D^+\rightarrow \overline{K}_0^*(1430)^0 e^+\nu_e$&$44^{+29}_{-19}$&$46^{+37}_{-26}$&$45.9^{+20.7}_{-16.7}$&$56.3^{+25.2}_{-20.4}$&&\\			
			$D^+\rightarrow \overline{K}_0^*(1430)^0 \mu^+\nu_\mu$&$32^{+25}_{-15}$&$46^{+37}_{-26}$&$35.2^{+16.0}_{-12.9}$&$43.1^{+19.5}_{-15.8}$&&$<23$\\		
			$D^0\rightarrow K_0^*(1430)^- e^+\nu_e$&$17^{+11}_{-8}$&$18^{+15}_{-10}$&$18.3^{+8.2}_{-6.7}$&$22.4^{+10.1}_{-8.1}$&&\\
			$D^0\rightarrow K_0^*(1430)^- \mu^+\nu_\mu$&$12^{+9}_{-6}$&$18^{+15}_{-10}$&$14.0^{+6.4}_{-5.1}$&$17.1^{+7.8}_{-6.3}$&&\\
			$D_s^+\rightarrow K_0^*(1430)^0e^+\nu_e$&$1.55^{+0.90}_{-0.61}$&$2.4^{+2.2}_{-1.5}$&&&$3.6^{+1.9}_{-1.4}$&\\
			$D_s^+\rightarrow K_0^*(1430)^0\mu^+\nu_\mu$&$1.27^{+0.81}_{-0.53}$&$2.4^{+2.2}_{-1.5}$&&&$3.1^{+1.6}_{-1.2}$&\\
\end{tabular}\label{BrCom2}
\end{ruledtabular}
\end{table}

In Tables~\ref{BrCom1}--\ref{BrCom3} we compare our results with the available experimental data \cite{ParticleDataGroup:2024cfk} and previous theoretical predictions \cite{Cheng2017,Qiao2024,Khosravi2009,Momeni2019,Yang2006,Yang2024,Huang2023,Huang2021,Qiao2024,Hu2022}. At present only the branching ratios of the $D^{+}\to \overline{K}_1(1270)^0e^+\nu_e$, $D^{+}\to \overline{K}_1(1270)^0\mu^+\nu_\mu$ and $D^0\to K_1(1270)^-e^+\nu_e$, $D^0\to K_1(1270)^-\mu^+\nu_\mu$  decays were measured experimentally. From Table~\ref{BrCom1} we see that our results as well as theoretical predictions of Refs.~\cite{Cheng2017,Qiao2024,Khosravi2009} are in a good agreement with experimental data. It should be noted  that predictions obtained in the framework of the light cone sum rules \citep{Momeni2019} almost an order of magnitude exceed the experimental values. For other branching fractions of the semileptonic charm meson decays to excited mesons only upper experimental bounds are available \cite{ParticleDataGroup:2024cfk}. Again for the $D_s^+\rightarrow K_1(1270)^0e^+\nu_e$ decay Ref.~\citep{Momeni2019} predicts the value which is about a factor of 4 higher than experimental limit, while our result and other theoretical predictions \cite{Cheng2017,Qiao2024,Khosravi2009}  are well consistent with this upper bound.  

 {\makeatletter
\renewcommand\table@hook{\footnotesize}
\makeatother
\begin{table}[hbt]
	\caption{Comparison of the calculated branching fractions with other theoretical predictions and experimental data for the semileptonic charm meson decays into orbitally excited light mesons  ($\times 10^{-5}$).}
	\begin{ruledtabular}
		\begin{tabular}{@{\!\!}c@{}c@{\!}ccc@{\!}c@{\!\!}c@{}c@{}}
			\text{Decay}&\text{Our}&\cite{Zuo2016}&\cite{Cheng2017}&\cite{Huang2021}&\cite{Qiao2024}&\cite{Hu2022}&\text{PDG}\cite{ParticleDataGroup:2024cfk}\\
			\hline
$D^+\rightarrow a_0(1450)^0 e^+\nu_e$ &$1.57^{+0.69}_{-0.53}$&&$0.54\pm0.05$&$0.428$&&& \\
$D^+\rightarrow a_0(1450)^0 \mu^+\nu_\mu$ &$1.13^{+0.58}_{-0.42}$&&$0.38\pm0.03$&$0.276$&&& \\
$D^0\rightarrow a_0(1450)^- e^+\nu_e$ &$1.18^{+0.51}_{-0.38}$&&&$0.314$&&& \\
$D^0\rightarrow a_0(1450)^- \mu^+\nu_\mu$ &$0.85^{+0.43}_{-0.32}$&&&$0.201$&&& \\
$D^+\rightarrow f_0(1500)^0 e^+\nu_e$ &$0.29^{+0.12}_{-0.10}$&&$0.11\pm0.02$&&&& \\
$D^+\rightarrow f_0(1500)^0 \mu^+\nu_\mu$ &$0.18^{+0.10}_{-0.07}$&&$0.07\pm0.01$&&&& \\
$D^+\rightarrow a_1(1260)^0e^+\nu_e$&$12.4^{+4.7}_{-3.6}$&$1.47^{+0.55}_{-0.44}$&&$9.38$&$5.79\pm3.00$&$6.673^{+0.947}_{-0.811}$&\\
$D^+\rightarrow a_1(1260)^0\mu^+\nu_\mu$&$10.5^{+4.2}_{-3.2}$&$1.47^{+0.55}_{-0.44}$&&$8.52$&$5.11\pm2.70$&$6.002^{+0.796}_{-0.748}$&\\
$D^0\rightarrow a_1(1260)^-e^+\nu_e$&$9.5^{+3.5}_{-2.7}$&$1.11^{+0.41}_{-0.34}$&&$6.9$&$4.46\pm2.32$&$5.261^{+0.745}_{-0.639}$&\\
$D^0\rightarrow a_1(1260)^-\mu^+\nu_\mu$&$8.0^{+3.1}_{-2.5}$&$1.11^{+0.41}_{-0.34}$&&$6.27$&$3.93\pm2.08$&$4.732^{+0.685}_{-0.590}$&\\
$D^+\rightarrow b_1(1235)^0e^+\nu_e$&$3.39\pm0.40$&&$7.4\pm0.7$&$6.58$&$3.41\pm1.88$&&$<17.5$\\
$D^+\rightarrow b_1(1235)^0\mu^+\nu_\mu$&$2.76\pm0.32$&&$6.4\pm0.6$&$6.00$&$2.99\pm1.65$&&\\
$D^0\rightarrow b_1(1235)^-e^+\nu_e$&$2.56\pm0.30$&&&$4.85$&$2.61\pm1.44$&&$<11.2$\\
$D^0\rightarrow b_1(1235)^-\mu^+\nu_\mu$&$2.08\pm0.24$&&&$4.40$&$2.29\pm1.26$&&\\
$D^+\rightarrow h_1(1415)^0e^+\nu_e$&$2.9\pm0.5(10^{-3})$&&$\{0,0.02\}$&&$0.11\pm0.11$&&\\
$D^+\rightarrow h_1(1415)^0\mu^+\nu_\mu$&$2.1\pm0.5(10^{-3})$&&$\{0,0.02\}$&&$0.08\pm0.08$&&\\
$D^+\rightarrow h_1(1170)^0e^+\nu_e$&$6.4\pm0.7$&&$14\pm1.50$&&$5.28\pm3$&&\\
$D^+\rightarrow h_1(1170)^0\mu^+\nu_\mu$&$5.4\pm0.6$&&$12.2\pm1.3$&&$4.73\pm2.69$&&\\
$D^+\to f_1(1285)^0e^+\nu_e$&$8.7\pm1.0$&$1.07^{+0.39}_{-0.33}$&$3.7\pm0.8$&&$1.88\pm1.88$&&\\
$D^+\to f_1(1285)^0\mu^+\nu_\mu$&$7.1\pm0.8$&$1.07^{+0.39}_{-0.33}$&$3.2\pm0.6$&&$1.61\pm1.61$&&\\
$D^+\to f_1(1420)^0e^+\nu_e$&$0.169\pm0.017$&$0.0122^{+0.0048}_{-0.0040}$&$\{0.02,0.14\}$&&$0.64\pm0.54$&&\\
$D^+\to f_1(1420)^0\mu^+\nu_\mu$&$0.126\pm0.013$&$0.0122^{+0.0048}_{-0.0040}$&$\{0.02,0.14\}$&&$0.48\pm0.41$&&\\
$D^+\to a_2(1320)^0e^+\nu_e$&$0.345\pm0.036$&&&&$0.47\pm0.17$&&\\
$D^+\to a_2(1320)^0\mu^+\nu_\mu$&$0.258\pm0.027$&&&&$0.36\pm0.12$&&\\
$D^0\to a_2(1320)^-e^+\nu_e$&$0.260\pm0.028$&&&&$0.35\pm0.13$&&\\
$D^0\to a_2(1320)^-\mu^+\nu_\mu$&$0.194\pm0.021$&&&&$0.27\pm0.09$&&\\
$D^+\to f_2(1270)^0e^+\nu_e$&$0.63\pm0.07$&&&&$0.78\pm0.29$&&\\
$D^+\to f_2(1270)^0\mu^+\nu_\mu$&$0.49\pm0.05$&&&&$0.62\pm0.21$&&\\
$D^+\to f'_2(1525)^0e^+\nu_e$&$2.6\pm0.2(10^{-4})$&&&&$5.35\pm2.83(10^{-4})$&&\\
$D^+\to f'_2(1525)^0\mu^+\nu_\mu$&$1.3\pm0.1(10^{-4})$&&&&$2.90\pm1.48(10^{-4})$&&\\
$D_s^+\rightarrow f_0(1500)^0e^+\nu_e$&$9.2^{+2.8}_{-2.3}$&&$15\pm3$&&&&\\
$D_s^+\rightarrow f_0(1500)^0\mu^+\nu_\mu$&$6.6^{+2.4}_{-1.8}$&&$12\pm2$&&&&\\
$D_s^+\rightarrow f_1(1285)^0e^+\nu_e$&$11.7\pm1.8$&&$\{6.0,36\}$&&$86\pm73$&&\\
$D_s^+\rightarrow f_1(1285)^0\mu^+\nu_\mu$&$10.1\pm1.2$&&$\{5.2,30.6\}$&&$76\pm65$&&\\
$D_s^+\rightarrow f_1(1420)^0e^+\nu_e$&$39.9\pm4.0$&&$25\pm5$&&$21\pm21$&&\\
$D_s^+\rightarrow f_1(1420)^0\mu^+\nu_\mu$&$32.1\pm3.3$&&$21\pm5$&&$18\pm18$&&\\
$D_s^+\rightarrow h_1(1415)^0e^+\nu_e$&$12.6\pm1.9$&&$64\pm7$&&$28\pm16$&&\\
$D_s^+\rightarrow h_1(1415)^0\mu^+\nu_\mu$&$9.8\pm1.5$&&$54\pm6$&&$24\pm14$&&\\
$D_s^+\rightarrow h_1(1170)^0e^+\nu_e$&$0.98\pm0.11$&&$\{0,19.7\}$&&$26\pm26$&&\\
$D_s^+\rightarrow h_1(1170)^0\mu^+\nu_\mu$&$0.85\pm0.10$&&$\{0,17.4\}$&&$24\pm24$&&\\
$D_s^+\to f_2(1270)^0e^+\nu_e$&$1.43\pm0.15$&&&&$1.04\pm0.54$&&\\
$D_s^+\to f_2(1270)^0\mu^+\nu_\mu$&$1.18\pm0.12$&&&&$0.87\pm0.44$&&\\
$D_s^+\to f_2'(1525)^0e^+\nu_e$&$2.90\pm0.32$&&&&$1.83\pm0.69$&&\\
$D_s^+\to f_2'(1525)^0\mu^+\nu_\mu$&$1.92\pm0.21$&&&&$1.25\pm0.45$&&\\

\end{tabular}\label{BrCom3}
\end{ruledtabular}
\end{table}
}

The results for the $D_{(s)}$ decays to the scalar strange meson $K^*_0(1430)$ are compared in Table~\ref{BrCom2}.  The central values of all listed theoretical predictions for the $D^{+}\to \overline{K}_0^*(1430)^0\mu^+\nu_\mu$ decay branching fraction, including ours, slightly exceed the experimental upper bound. However, all calculated values have large uncertainties, which mainly originate from the poorly measured ${K}_0^*(1430)$ mass. As a result they are compatible with experimental limit \cite{ParticleDataGroup:2024cfk} within the errorbars \cite{Yang2006,Yang2024}.  The closeness of the predictions to the experimental upper bound may be a sign of an experimental detection of this decay in a near future. Note that all listed theoretical predictions for the charmed meson decays to the scalar $K^*_0(1430)$ have close values consistent within uncertainties. 

Predictions for the semileptonic charm meson decays to excited light mesons are compared in Table~\ref{BrCom3}. We find that our results reasonably agree with calculations in Refs.~\cite{Qiao2024,Huang2021,Hu2022,Cheng2017}. However, there is almost an order of magnitude difference for the $D\to a_1(1260)l\nu_l$ decays with the 3-point QCD sum rule values \cite{Zuo2016} of branching fractions. The predicted values of the $D\to b_1(1235)e\nu_e$ branching fractions are consistent with the upper bound form PDG \cite{ParticleDataGroup:2024cfk}. Very recently the BESIII Collaboration reported observation of the $D^0\to b_1(1235)^-e^+\nu_e$ and evidence for the $D^+\to b_1(1235)^0e^+\nu_e$ decay \citep{BESIII:2024pwp}. The following products of branching fractions were determined
\begin{eqnarray}\label{Db1}
Br(D^0 \to b_1(1235)^-e^+\nu_e) \times Br(b_1(1235)^- \to \omega\pi^-) &=& (0.72 \pm 0.18_{-0.08}^ {+0.06}) \times 10^{-4}, \cr \cr
 Br(D^+ \to b_1(1235)^0 e^+ \nu_e) \times Br(b_1(1235)^0 \to \omega\pi^0 )& =& (1.16 \pm 0.44 \pm 0.16) \times 10^{-4}.
\end{eqnarray}
These data are somewhat higher than our results given in Table \ref{BrCom3}  but they are consistent within $2\sigma$.

It is important to point our that in this paper we present the first (to our knowledge) detailed dynamical study of the semileptonic $D_{(s)}$ decays to tensor strange and light mesons as well as radially excited mesons. Previously $D_{(s)}$ decays to tensor mesons were considered only on the basis of the SU(3) flavor symmetry \citep{Qiao2024}. Let us also point out that our results for all considered semileptonic  $D$ decays  into orbitally excited strange and light mesons are well consistent with SU(3) flavor symmetry predictions \citep{Qiao2024}. Our value for the branching fraction of the semileptonic $D$ decay to the radially excited $K^*(1680)$ meson Br$(D^+\rightarrow \overline{K}^*(1680)^0e^+\nu_e)=0.33^{+0.24}_{-0.15}\times 10^{-5}$ (see Table~\ref{Br2}) is well below experimental upper bound Br$(D^+\rightarrow \overline{K}^*(1680)^0e^+\nu_e)<1.5\times 10^{-3}$ \cite{ParticleDataGroup:2024cfk}. 

Recently possible hints of the violation of the lepton universality were found in $B$ decays, where deviations from the standard model predictions for the ratios of the semileptonic decay branching fractions involving $\tau$ lepton and electron were reported (for a review see, e.g., Ref.\cite{Bernlochner:2021vlv}). Semileptonic decays of $D$ mesons involving  the $\tau$ lepton are forbidden  by the energy conservation due to the high value of the $\tau$ mass. Thus for $D$ decays it is reasonable to check lepton universality comparing semileptonic decays with $\mu$ and positron.  In Table~\ref{Ratio} we give our results for the corresponding ratios of the $D$ semileptonic decays
\begin{equation}
R_F=\dfrac{\Gamma(D_{(s)}\to F\mu^+\nu_\mu)}{\Gamma(D_{(s)}\to Fe^+\nu_e)},
\end{equation}
where $F$ is the excited strange or light meson. Note that almost all theoretical uncertainties cancel in these ratios. If future experimental measurements find significant deviations from the presented values, this will signal the presence of the so-called new physics beyond the standard model.

\begin{table}[hbt]
	\caption{Ratio of the branching fractions with $\mu$ and $e$.}
	\begin{ruledtabular}
		\begin{tabular}{l c|l c}
			\text{Decay}&\text{$R_F$}&\text{Decay}&\text{$R_F$}\\	
			\hline
			$D\rightarrow K_0^*(1430)l^+\nu_l$&0.727&	$D_s\rightarrow K_0^*(1430)l^+\nu_l$&0.819\\
			$D\rightarrow K_1(1270)l^+\nu_l$&0.899&$D_s\rightarrow K_1(1270)l^+\nu_l$&0.903\\
			$D\rightarrow K_1(1400)l^+\nu_l$&0.730&$D_s\rightarrow K_1(1400)l^+\nu_l$&0.794\\
			$D\rightarrow K_2^*(1430)l^+\nu_l$&0.641&$D_s\rightarrow K_2^*(1430)l^+\nu_l$&0.745\\
			$D\rightarrow a_0(1450)l^+\nu_l$&0.720&$D_s\rightarrow f_0(1500)l^+\nu_l$&0.717\\
			$D\rightarrow a_1(1260)l^+\nu_l$&0.847&$D_s\rightarrow f_1(1420)l^+\nu_l$&0.805\\
			$D\rightarrow b_1(1235)l^+\nu_l$&0.814&$D_s\rightarrow h_1(1415)l^+\nu_l$&0.778\\
			$D\rightarrow h_1(1170)l^+\nu_l$&0.844&$D_s\rightarrow h_1(1170)l^+\nu_l$&0.867\\
			$D\rightarrow h_1(1415)l^+\nu_l$&0.707&$D_s\rightarrow f_2(1270)l^+\nu_l$&0.825\\
			$D\rightarrow a_2(1320)l^+\nu_l$&0.748&$D_s\rightarrow f_2'(1525)l^+\nu_l$&0.663\\
			$D\to f_0(1370)l^+\nu_l$&0.802&$D_s\rightarrow f_1(1285)l^+\nu_l$&0.863\\
			$D\to f_0(1500)l^+\nu_l$&0.620&$D_s\rightarrow f_0(1370)l^+\nu_l$&0.841\\
			$D\to f_1(1285)l^+\nu_l$&0.816&$D\to f_2(1270)l^+\nu_l$&0.778\\
			$D\to f_1(1420)l^+\nu_l$&0.746&$D\to f_2'(1525)l^+\nu_l$&0.517\\
			$D\rightarrow K(1460)l^+\nu_l$&0.646&$D_s\rightarrow K(1460)l^+\nu_l$&0.785\\
			$D\rightarrow K^*(1680)l^+\nu_l$&0.067&$D_s\rightarrow K^*(1680)l^+\nu_l$&0.419\\
			$D\rightarrow \pi(1300)l^+\nu_l$&0.820&$D_s\rightarrow \eta(1475)l^+\nu_l$&0.777\\
			$D\rightarrow \rho(1450)l^+\nu_l$&0.733&$D_s\rightarrow \phi(1680)l^+\nu_l$&0.543\\
			$D\rightarrow \eta(1295)l^+\nu_l$&0.825&$D\rightarrow \omega(1420)l^+\nu_l$&0.783\\	
			
		\end{tabular}\label{Ratio}
	\end{ruledtabular}
\end{table}

\begin{figure}
	\begin{minipage}[h]{0.47\linewidth}
		\center{\includegraphics[width=1\linewidth]{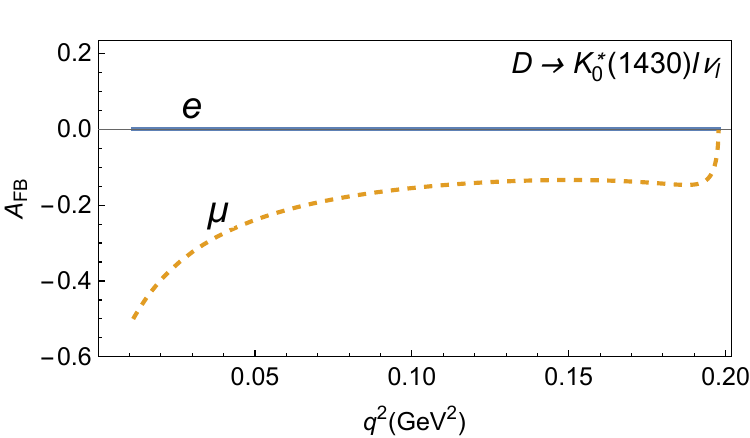}} \\
	\end{minipage}
	\hfill
	\begin{minipage}[h]{0.47\linewidth}
		\center{\includegraphics[width=1\linewidth]{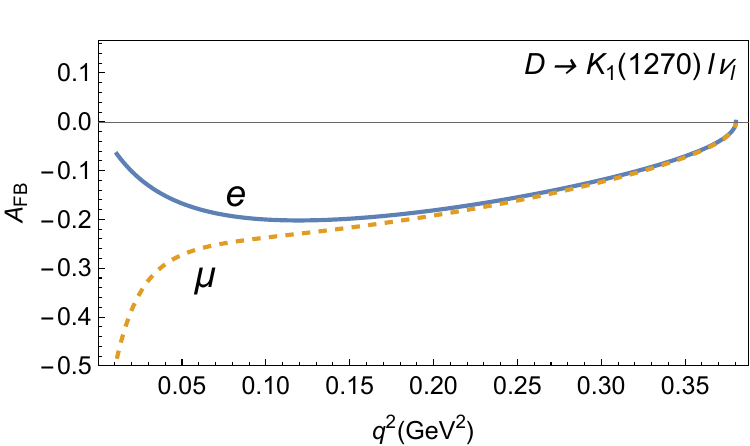}} \\
	\end{minipage}
	\vfill
	\begin{minipage}[h]{0.47\linewidth}
		\center{\includegraphics[width=1\linewidth]{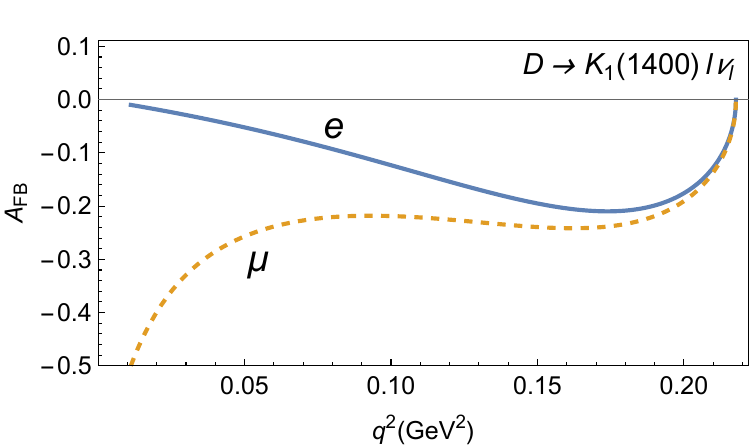}} \\
	\end{minipage}
	\hfill
	\begin{minipage}[h]{0.47\linewidth}
		\center{\includegraphics[width=1\linewidth]{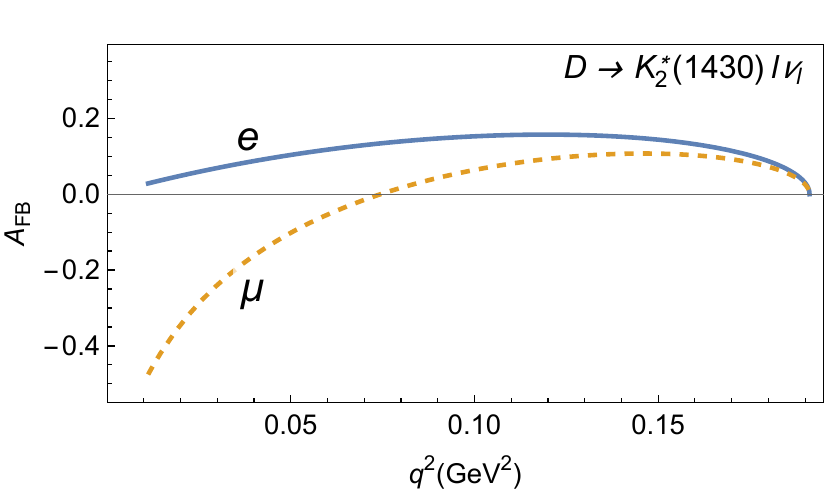}} \\
	\end{minipage}
	\caption{Forward-backward asymmetry $A_{FB}$ for the semileptonic $D$ decays into orbitally excited kaons.}
	\label{GrAfb}
\end{figure}

\begin{figure}
	\begin{minipage}[h]{0.47\linewidth}
		\center{\includegraphics[width=1\linewidth]{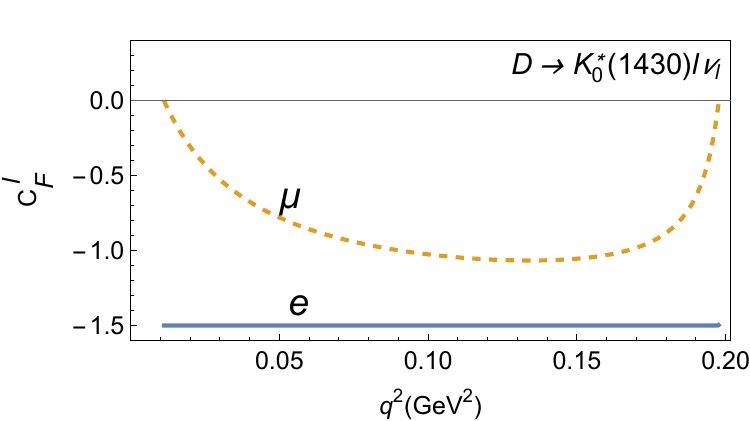}} \\
	\end{minipage}
	\hfill
	\begin{minipage}[h]{0.47\linewidth}
		\center{\includegraphics[width=1\linewidth]{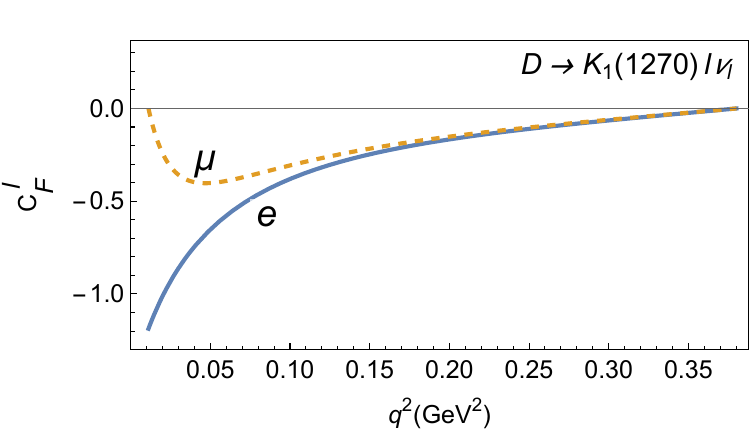}} \\
	\end{minipage}
	\vfill
	\begin{minipage}[h]{0.47\linewidth}
		\center{\includegraphics[width=1\linewidth]{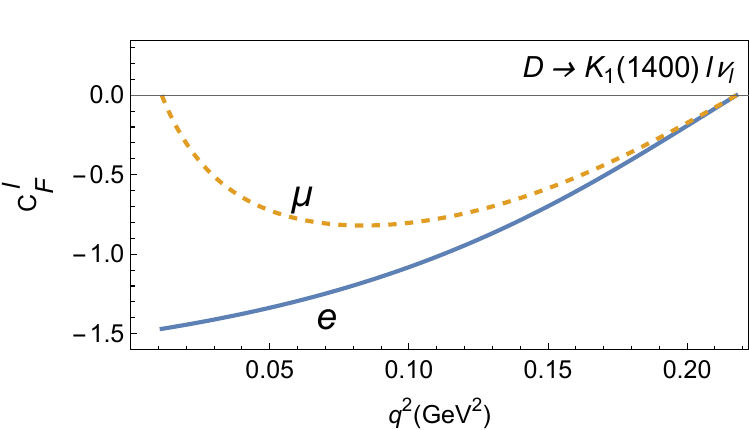}} \\
	\end{minipage}
	\hfill
	\begin{minipage}[h]{0.47\linewidth}
		\center{\includegraphics[width=1\linewidth]{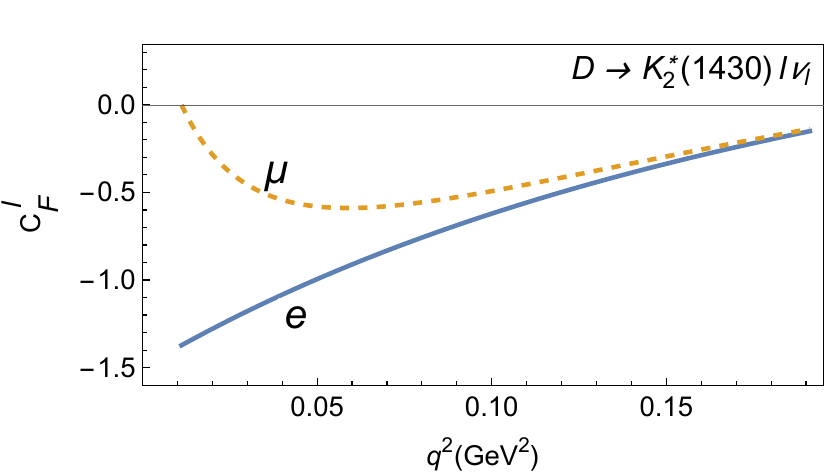}} \\
	\end{minipage}
	\caption{Lepton-side convexity parameter $C^l_{F}$ for the semileptonic $D$ decays into orbitally excited kaons.}
	\label{GrClf}
\end{figure}

To complete our analysis of the semileptonic charm meson decays to the excited strange and light mesons  we calculate the forward-backward asymmetry $A_{FB}(q^2)$ defined in (\ref{Afb}), the lepton-side convexity parameter $C^l_F(q^2)$ (\ref{Clf}), the longitudinal $P^l_L(q^2)$ (\ref{Pll}) and transverse
$P^l_T(q^2)$ (\ref{Plt}) polarizations of the final charged lepton, and longitudinal polarization $F_L(q^2)$ (\ref{Fl}) of the final vector meson.  As an example in Figs.~\ref{GrAfb},~\ref{GrClf} and  we plot the forward-backward asymmetry $A_{FB}(q^2)$ and the lepton-side convexity parameter $C^l_F(q^2)$ for the $D^+\to K^{(*)}_{0,1,2}l\nu_l$ decays.

In Tables~\ref{Ob1}--\ref{Ob3} we present our predictions for the mean values of the polarization and asymmetry parameters for the semileptonic $D$ and $D_s$ decays. These values were obtained by separately integrating corresponding partial differential decay rates in numerators and the total decay rates in denominators. Since we neglect the small positron mass, for all decays $D^+_{s}\to S(P)e^+\nu_e,$ where $S$ and $P$ are scalar and pseudoscalar mesons, respectively,  $\langle P^e_L\rangle=1$ and $\langle P^e_T\rangle=0$, while for decays $D^+_{s}\to Fe^+\nu_e,$ $\langle A_{FB}\rangle=0$ and $\langle C^e_L\rangle=-1.5$. If the small positron mass is taken into account only tiny deviations from these values arise.

\begin{table}[hbt]
	\caption{Predictions for the asymmetry and polarization parameters for the semileptonic $D$ decays into orbitally excited strange and light mesons with the positively charged leptons.}
	\begin{ruledtabular}
		\begin{tabular}{c c c c c c}
			\text{Decay}& $\langle A_{FB}\rangle $& $\langle C^l_F\rangle $ &$\langle P_L\rangle $&$\langle P_T\rangle $&$\langle F_L\rangle $\\
			\hline
			$D\rightarrow K_0^*(1430)e\nu_e$&0 &-1.5&1 &0 &--\\
			$D\rightarrow K_0^*(1430)\mu\nu_\mu$&-0.213 &-0.850 &0.386 &-0.776 &--\\
			$D\rightarrow K_1(1270)e\nu_e$&-0.148 &-0.199 &1 &0 &0.422\\
			$D\rightarrow K_1(1270)\mu\nu_\mu$&-0.165 &-0.140 &0.908 &-0.067 &0.417\\
			$D\rightarrow K_1(1400)e\nu_e$&-0.085 &-1.166 &1 &0 &0.851\\
			$D\rightarrow K_1(1400)\mu\nu_\mu$&-0.250 &-0.654 &0.553 &-0.531 &0.830\\
			$D\rightarrow K_2^*(1430)e\nu_e$&0.096 &-0.964 &1 &0 &0.762\\
			$D\rightarrow K_2^*(1430)e\nu_e$&-0.048 &-0.489 &0.589 &-0.541 &0.729\\
			$D\rightarrow a_0(1450)e\nu_e$&0 &-1.5 &1 &0 &-- \\
			$D\rightarrow a_0(1450)\mu\nu_\mu$&-0.223 &-0.814 &0.343 &-0.797 &--\\
			$D\rightarrow a_1(1260)e\nu_e$&-0.111 &-0.877&1 &0 &0.723\\
			$D\rightarrow a_1(1260)\mu\nu_\mu$&-0.198 &-0.599 &0.763 &-0.327 &0.706\\
			$D\rightarrow b_1(1235)e\nu_e$&0.026&-1.074&1&0&0.811\\
			$D\rightarrow b_1(1235)\mu\nu_\mu$&-0.125&-0.561&0.522&-0.611&0.790\\
			$D\rightarrow h_1(1170)e\nu_e$&0.022&-1.071&1&0&0.809\\
			$D\rightarrow h_1(1170)\mu\nu_\mu$&-0.117&-0.603&0.565&-0.591&0.792\\
			$D\rightarrow h_1(1415)e\nu_e$&0.061&-0.850&1&0&0.711\\
			$D\rightarrow h_1(1415)\mu\nu_\mu$&-0.103&-0.211&0.393&-0.572&0.661\\
			$D\rightarrow a_2(1320)e\nu_e$&0.137&-0.975&1&0&0.767\\
			$D\rightarrow a_2(1320)\mu\nu_\mu$&0.015&-0.575&0.649&-0.538&0.745\\
			$D\to f_0(1370)e\nu_e$&0&-1.5&1&0&--\\
			$D\to f_0(1370)\mu\nu_\mu$&-0.182&-0.946&0.479&-0.732&--\\
			$D\to f_0(1500)e\nu_e$&0&-1.5&1&0&--\\
			$D\to f_0(1500)\mu\nu_\mu$&-0.266&-0.675&0.197&-0.843&--\\
			$D\to f_1(1285)e\nu_e$&-0.053&-1.146&1&0&0.843\\
			$D\to f_1(1285)\mu\nu_\mu$&-0.182&-0.746&0.644&-0.508&0.830\\
			$D\to f_1(1420)e\nu_e$&-0.134&-0.572&1&0&0.588\\
			$D\to f_1(1420)\mu\nu_\mu$&-0.213&-0.301&0.749&-0.226&0.558\\
			$D\to f_2(1270)e\nu_e$&0.132&-0.992&1&0&0.774\\
			$D\to f_2(1270)\mu\nu_\mu$&0.016&-0.614&0.668&-0.526&0.755\\
			$D\to f_2^{'}(1525)e\nu_e$&0.162&-0.879&1&0&0.724\\
			$D\to f_2^{'}(1525)\mu\nu_\mu$&0.006&-0.356&0.538&-0.585&0.677\\
			\end{tabular}\label{Ob1}
	\end{ruledtabular}
\end{table}

\begin{table}[hbt]
	\caption{Predictions for the asymmetry and polarization parameters for the semileptonic $D_s$ decays into orbitally excited strange and light mesons with the positively charged leptons.}
	\begin{ruledtabular}
		\begin{tabular}{c c c c c c}
			\text{Decay}& $\langle A_{FB}\rangle $& $\langle C^l_F\rangle $ &$\langle P_L\rangle $&$\langle P_T\rangle $&$\langle F_L\rangle $\\
			\hline
			$D_s\rightarrow K_0^*(1430)e\nu_e$&0 &-1.5&1 &0 &--\\ 
			$D_s\rightarrow K_0^*(1430)\mu\nu_\mu$&-0.156 &-1.030 &0.577 &-0.659 &--\\
			$D_s\rightarrow K_1(1270)e\nu_e$&-0.200 &-0.319 &1 &0 &0.422\\
			$D_s\rightarrow K_1(1270)\mu\nu_\mu$&-0.234 &-0.204 &0.885 &-0.096 &0.462\\
			$D_s\rightarrow K_1(1400)e\nu_e$&-0.106 &-1.054 &1 &0 &0.802\\
			$D_s\rightarrow K_1(1400)\mu\nu_\mu$&-0.227 &-0.678 &0.678 &-0.423 &0.785\\
			$D_s\rightarrow K_2^*(1430)e\nu_e$&0.137 &-0.929 &1 &0 &0.746\\
			$D_s\rightarrow K_2^*(1430)e\nu_e$&0.018 &-0.544 &0.661 &-0.517 &0.724\\
			$D_s\rightarrow f_0(1370)e\nu_e$&0 &-1.5 &1 &0 &-- \\
			$D_s\rightarrow f_0(1370)\mu\nu_\mu$&-0.123 &-1.128 &0.683 &-0.556 &--\\
			$D_s\rightarrow f_0(1500)e\nu_e$&0 &-1.5 &1 &0 &-- \\
			$D_s\rightarrow f_0(1500)\mu\nu_\mu$&-0.195 &-0.911 &0.473 &-0.715 &--\\
			$D_s\rightarrow f_1(1285)e\nu_e$&-0.218 &-0.723&1 &0 &0.655\\
			$D_s\rightarrow f_1(1285)\mu\nu_\mu$&-0.290 &-0.494 &0.806 &-0.223 &0.637\\
			$D_s\rightarrow f_1(1420)e\nu_e$&-0.223 &-0.649&1 &0 &0.620\\
			$D_s\rightarrow f_1(1420)\mu\nu_\mu$&-0.303 &-0.387 &0.769 &-0.210 &0.598\\
			$D_s\rightarrow h_1(1170)e\nu_e$&0&-1.5&1&0&0.989\\
			$D_s\rightarrow h_1(1170)\mu\nu_\mu$&-0.132&-1.060&0.613&-0.642&0.988\\
			$D_s\rightarrow h_1(1415)e\nu_e$&0.024&-0.876&1&0&0.723\\
			$D_s\rightarrow h_1(1415)\mu\nu_\mu$&-0.127&-0.331&0.489&-0.556&0.688\\
			$D_s\rightarrow f_2(1270)e\nu_e$&0.080&-1.026&1&0&0.789\\
			$D_s\rightarrow f_2(1270)\mu\nu_\mu$&-0.023&-0.698&0.715&-0.473&0.775\\
			
			$D_s\rightarrow f_2'(1525)e\nu_e$&0.096&-0.944&1&0&0.753\\
			$D_s\rightarrow f_2'(1525)\mu\nu_\mu$&-0.042&-0.491&0.610&-0.525&0.722\\
			\end{tabular}\label{Ob2}
	\end{ruledtabular}
\end{table}
			\begin{table}[hbt]
	\caption{Predictions for the asymmetry and polarization parameters for the semileptonic charm meson decays into radially excited strange and light mesons with the positively charged leptons.}
	\begin{ruledtabular}
		\begin{tabular}{c c c c c c}
		\text{Decay}& $\langle A_{FB}\rangle $& $\langle C^l_F\rangle $ &$\langle P_L\rangle $&$\langle P_T\rangle $&$\langle F_L\rangle $\\
			\hline
			$D\rightarrow K(1460)e\nu_e$&0&-1.5&1&0&--\\
			$D\rightarrow K(1460)\mu\nu_\mu$&-0.230&-0.806&0.366&-0.768&--\\
			$D\rightarrow K^*(1680)e\nu_e$&-0.081&-0.497&1&0&0.554\\
			$D\rightarrow K^*(1680)\mu\nu_\mu$&-0.171&-0.027&0.441&-0.163&0.421\\
			$D\rightarrow \eta(1295)e\nu_e$&0&-1.5&1&0&--\\
			$D\rightarrow \eta(1295)\mu\nu_\mu$&-0.155&-1.033&0.574&-0.660&--\\
			$D\rightarrow \pi(1300)e\nu_e$&0&-1.5&1&0&--\\
			$D\rightarrow \pi(1300)\mu\nu_\mu$&-0.158&-1.025&0.566&-0.666&--\\
			$D\rightarrow \omega(1420)e\nu_e$&-0.234&-0.393&1&0&0.508\\
			$D\rightarrow \omega(1420)\mu\nu_\mu$&-0.289&-0.203&0.788&-0.130&0.489\\
			$D\rightarrow \rho(1450)e\nu_e$&-0.218&-0.415&1&0&0.518\\
			$D\rightarrow \rho(1450)\mu\nu_\mu$&-0.282&-0.193&0.753&-0.142&0.494\\
			$D_s\rightarrow K(1460)e\nu_e$&0&-1.5&1&0&--\\
			$D_s\rightarrow K(1460)\mu\nu_\mu$&-0.177&-0.966&0.513&-0.699&--\\
			$D_s\rightarrow K^*(1680)e\nu_e$&-0.098&-0.505&1&0&0.558\\
			$D_s\rightarrow K^*(1680)\mu\nu_\mu$&0.200&-0.127&0.596&-0.226&0.495\\
			$D_s\rightarrow \eta(1475)e\nu_e$&0&-1.5&1&0&--\\
			$D_s\rightarrow \eta(1475)\mu\nu_\mu$&-0.185&-0.939&0.481&-0.724&--\\
			$D_s\rightarrow \phi(1680)e\nu_e$&-0.114&-0.490&1&0&0.551\\
			$D_s\rightarrow \phi(1680)\mu\nu_\mu$&-0.208&-0.158&0.647&-0.216&0.503\\
		\end{tabular}\label{Ob3}
	\end{ruledtabular}
\end{table}

\section{Conclusion}
\label{sec:concl}

Exclusive semileptonic decays of  $D$ and $D_s$ meson into orbitally and radially excited kaons and light mesons were studied in the framework of the relativistic quark model. The hadronic matrix elements of the weak current between the $D_{(s)}$ mesons and orbitally excited scalar, axial-vector and tensor as well as radially excited pseudoscalar and vector mesons were calculated with the consistent account of relativistic effects using the quasipotential approach. On this basis the invariant form factors, which parameterize these matrix elements, were obtained as the overlap integrals of the corresponding meson wave functions. These wave functions are known from the previous calculations of the meson mass spectra  \cite{Ebert2009,Ebert2010a}. The mixing between $q\bar q$ ($q=u,d$) and $s\bar s$ states in the isoscalar sector and of the axial-vector spin-singlet ($^1P_1$) and spin-triplet ($^3P_1$) states of kaons was discussed and taken into account in  calculations of the form factors. As a result the weak transition form factors were obtained in the whole accessible kinematical range without additional model assumptions and extrapolations. This is a significant advantage of our model since in most of the other theoretical approaches the form factors are calculated at a single value of  the transferred momentum squared $q^2$ (usually at the zero recoil point $q^2=0$), and then extrapolated to the whole kinematical range employing some model parameterizations.    
It was found that our numerical results for the form factors and their $q^2$ dependence can be accurately approximated by Eqs.~(\ref{Ap1})--(\ref{Ap3}). The
parameters of the fit are collected in Tables~\ref{FF1}--\ref{FF5}. Note that these form factors can be used for calculations not only of semileptonic decays but also for the nonleptonic decays in the factorization approximation \cite{Yu:2022ngu}.

These form factors were applied for the calculation of the differential and total decay rates of the semileptonic decays of $D_{(s)}$ mesons using the helicity formalism. The detailed comparison of the obtained results with the previous calculations  \cite{Cheng2017,Qiao2024,Khosravi2009,Momeni2019,Yang2006,Zuo2016,Yang2024,Huang2023,Huang2021,Qiao2024,Hu2022} and available experimental data \cite{ParticleDataGroup:2024cfk} is presented in Tables~\ref{BrCom1}--\ref{BrCom3}. Good agreement of our predictions with the measured $D\to \overline{K}_1(1270)e^+\nu_e$, $D\to \overline{K}_1(1270)\mu^+\nu_\mu$ decay branching fractions and with the upper bounds available for some decay modes is obtained.  The predicted $D^{+}\to \overline{K}_0^*(1430)^0\mu^+\nu_\mu$ decay branching fraction has rather large uncertainties and coincides within them with the experimental upper limit, indicating that this decay has good chances to be observed in near future. A very recent BESIII data on the semileptonic  $D\to b_1(1235)e^+\nu_e$ decays \citep{BESIII:2024pwp} is also compatible with our results. Note that our predictions for the semileptonic $D_{(s)}$ decays to orbitally excited kaons and light mesons agree well with the results obtained on the basis of the SU(3) flavor symmetry \citep{Qiao2024}. Let us point out that we performed the first (to our knowledge) dynamical calculation of the semileptonic $D_{(s)}$ decays to the orbitally excited tensor  and radially excited pseudoscalar and vector mesons. Previously branching fractions for decays to tensor mesons were obtained only in the SU(3) flavor symmetry framework~\citep{Qiao2024}.

In conclusion we note that the further increase of the experimental accuracy and new measurements can help to better understand quark dynamics in mesons. For this purpose it will be important not only to measure the total decay branching fractions, but also to study different differential decay distributions determining asymmetry and polarization parameters (e.g., the ones given in Tables~\ref{Ob1}--\ref{Ob3}).   Such measurements could be made, e.g., at the future super charm-tau factory \cite{Charm-TauFactory:2013cnj,Barnyakov_2020,Peng:2020orp,Cheng:2022tog}.

\begin{acknowledgments}
	The authors are grateful to  A.V. Berezhnoy for useful discussions. The work of Ivan S. Sukhanov was supported in part by the Foundation for the Advancement of Theoretical Physics and Mathematics ``BASIS'' grant number 23-2-2-11-1.
\end{acknowledgments}

\bibliography{Semileptonic}

\end{document}